\documentclass[journal,twocolumn]{IEEEtran}
\usepackage{amsmath}

\usepackage{amsthm}
\usepackage{exscale,relsize}
\usepackage{cite}
\usepackage{graphicx}
\usepackage{cases}
\usepackage{epstopdf}
\usepackage{amsfonts,amsmath,amssymb}
\usepackage{graphicx}
\usepackage{bm}
\usepackage{bbm}
\usepackage{color}
\usepackage{caption}
\usepackage{enumerate}

\theoremstyle{definition}

\usepackage{subfigure}
\renewcommand{\qedsymbol}{\square}
\usepackage[ruled,vlined,lined,ruled,linesnumbered]{algorithm2e}
\ifCLASSINFOpdf
\else
\fi
\begin{document}
\bstctlcite{IEEEexample:BSTcontrol}
%\title{Bare Demo of IEEEtran.cls\\ for IEEE Journals}
\title{{Capacity Analysis and Throughput Maximization of NOMA with Nonlinear Power Amplifier Distortion}

\author{Xiaojia~Wang,~\IEEEmembership{Graduate Student Member,~IEEE},
	Suzhi~Bi,~\IEEEmembership{Senior Member,~IEEE},
	Xian~Li,~\IEEEmembership{Member,~IEEE},
	Xiaohui~Lin,
	Zhi~Quan,~\IEEEmembership{Senior Member,~IEEE},
    and Ying-Jun~Angela~Zhang,~\IEEEmembership{Fellow,~IEEE}}

\thanks{This article was presented in part at the IEEE International Conference on Communications, Rome, Italy, May 28 to June 1, 2023 \cite{wangCapacity}. X.~Wang, S.~Bi, X.~Li, X.~Lin, and Z. Quan are with the College of Electronics and Information Engineering, Shenzhen University, China (wangxiaojia2021@email.szu.edu.cn, $\left\{\text{bsz,xianli,xhlin,zquan}\right\}$@szu.edu.cn). Y.-J. A. Zhang is with Department of Information Engineering, The Chinese University of Hong Kong, Hong Kong SAR (yjzhang@ie.cuhk.edu.hk). (\emph{Corresponding author: Suzhi Bi})}}
%	Part of this work has been presented in IEEE International Conference on Communications (ICC) 2023 \cite{LiICCC2021}.}}


%\author{Michael~Shell,~\IEEEmembership{Member,~IEEE,}
%        John~Doe,~\IEEEmembership{Fellow,~OSA,}
%        and~Jane~Doe,~\IEEEmembership{Life~Fellow,~IEEE}% <-this % stops a space
% <-this % stops a space
\maketitle

\begin{abstract}
In future B5G/6G broadband communication systems, non-linear signal distortion caused by the impairment of transmit power amplifier (PA) can severely degrade the communication performance, especially when uplink users share the wireless medium using non-orthogonal multiple access (NOMA) schemes.
This is because the successive interference cancellation (SIC) decoding technique, used in NOMA, is incapable of eliminating the interference caused by PA distortion. Consequently, each user's decoding process suffers from the cumulative distortion noise of all uplink users.
In this paper, we establish a new and tractable PA distortion signal model based on real-world measurements, where the distortion noise power is a polynomial function of PA transmit power diverging from the oversimplified linear function commonly employed in existing studies.
Applying the proposed signal model, we characterize the capacity rate region of multi-user uplink NOMA by optimizing the user transmit power. Our findings reveal a significant contraction in the capacity region of NOMA, attributable to polynomial distortion noise power.
For practical engineering applications, we formulate a general weighted sum rate maximization (WSRMax) problem under individual user rate constraints. We further propose an efficient power control algorithm to attain the optimal performance.
Numerical results show that the optimal power control policy under the proposed non-linear PA model achieves on average 13\% higher throughput compared to the policies assuming an ideal linear PA model.
Overall, our findings demonstrate the importance of accurate PA distortion modeling to the performance of NOMA and provide efficient optimal power control method accordingly.

\end{abstract}
\vspace{-5pt}
\begin{IEEEkeywords}\vspace{-0em}
	NOMA, power control, nonlinear power amplifier, capacity region.
\end{IEEEkeywords}

\section{Introduction}
\subsection{Motivations and Contributions}
{\color{black} Non-orthogonal multiple access (NOMA) has been widely recognized as a key enabling technology for B5G/6G wireless communication networks due to its high spectral efficiency \cite{dingSurvey2017,islamPowerDomain2017,vaeziNOMA2019,tangAchievable2019}.
Via successive interference cancellation (SIC) information decoding technique, NOMA allows uplink user signals to be superimposed in the same time-frequency resource block and sequentially decoded in the power domain. As a result, NOMA achieves a larger capacity region than its orthogonal multiple access (OMA) counterpart, providing benefits to all the participating users.
	\begin{figure}[tbp]
		\centering
		\vspace{2pt}
		\includegraphics[scale=0.6]{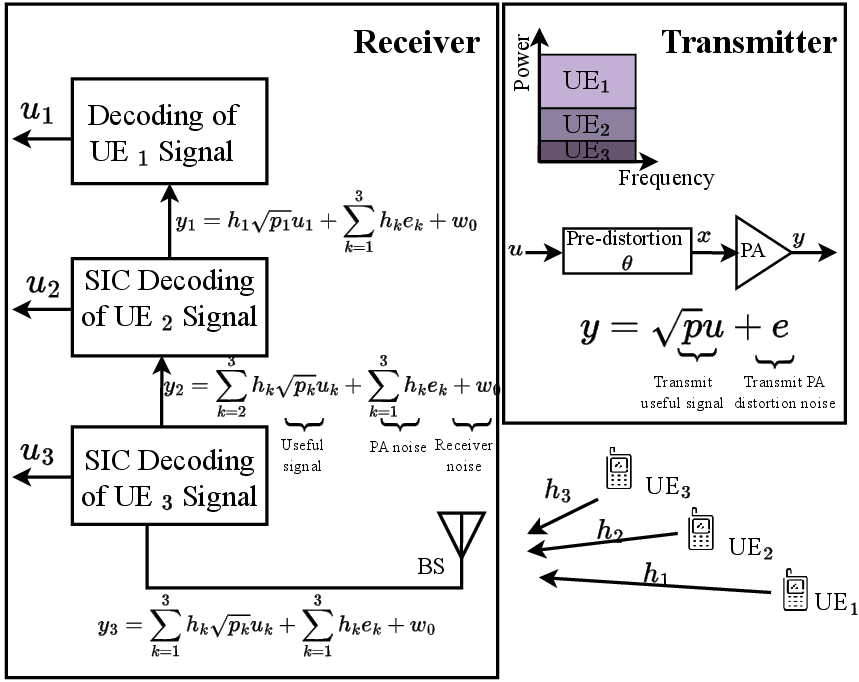}
		\captionsetup{font=footnotesize}
		\caption{Uplink NOMA system model under PA non-linear distortion.}
		\label{SIC}
	\end{figure}
B5G/6G wireless networks operate in a wide communication bandwidth to meet the soaring data rate demands. However, wideband orthogonal frequency division multiplexing (OFDM) modulation is known to suffer from a large peak-to-average power ratio (PAPR) of the transmitted signal \cite{benmabroukPrecodingbased2017}. As a result, the transmitted OFDM signals frequently enter the saturation operating region of the power amplifier (PA), leading to severe non-linear distortion of the baseband signals \cite{jiangOverview2008}. The nonlinear PA impairment can generate both in-band and out-of-band distortion that reduces the communication rate and creates interference on adjacent channels \cite{gregorioAnalysis}. To mitigate the distortion and attain higher PA efficiency, digital predistortion (DPD) is widely adopted to compensate for the nonlinear behavior of PA in the digital baseband domain \cite{chani-cahuanaDigital}. As shown in Fig. \ref{fig:DPD_a}, DPD pre-distorts the amplitude and phase of the PA input signal without the need for complex RF circuit design. This ensures that the output of the concatenated DPD-PA system linearly amplifies the input.
However, high-precision DPD is expensive and energy-consuming in operation, and residual distortion may persist even after DPD application \cite{bjornsonHardware2013}. Although NOMA enjoys higher spectral efficiency than conventional OMA schemes, it is more susceptible to nonlinear PA distortion. As shown in Fig. \ref{SIC}, the SIC procedure cannot fully eliminate the distortion noise, causing the decoding operation of each user to suffer from the cumulative distortion noise of all uplink users.

The impact of Power Amplifier (PA) distortion has been the focus of significant research recently \cite{bjornsonHardware2013,moghadamEnergy2018,korpiFullDuplex2014}.
%,costaImpact1999,dardariTheoretical2000,aggarwalEndtoend2019,qiPower2012,liResidual2020,guerreiroReceiver2020
\cite{bjornsonHardware2013} reveals that the nonlinear PA effect imposes a performance limit on the estimation accuracy and communication capacity of large-scale MISO systems. \cite{moghadamEnergy2018} shows that under nonlinear PA distortion, escalating the transmit power level can be counter-productive in terms of spectral and energy efficiency in a hybrid beamforming. This study also designs specialized beamforming filters for maximizing energy efficiency.
It is worth noting that the above works, among many others, model the PA distortion as an independent additive noise with power proportional to transmit power \cite{bjornsonHardware2013,korpiFullDuplex2014}. This simplified model, however, cannot accurately capture the common memory effect of the PA distortion, let alone the fact that the PA distortion behavior relates to many factors, such as the transmit power and bandwidth. Intuitively, the ratio between the noise power and PA transmit power is marginal when the transmit power is small. However, the ratio drastically escalates when the transmit power enters the PA saturation operating region. In addition, none of the existing models account for the residual distortion behavior of PA after DPD is used.

In this paper, we provide a new and analytically tractable PA non-linear distortion model based on real-world measurements. Based on the new signal model, we characterize the capacity region of the general NOMA uplink channel with PA non-linear distortion. For practical engineering applications, we propose an efficient power control algorithm to maximize the weighted sum rate of NOMA uplink users.
The main contributions of this paper are:
\vspace{-4pt}
\begin{itemize}
	\item \emph{Tractable signal model of nonlinear PA distortion:} We propose a new and tractable PA distortion model based on real-world measurements, where the distortion power is a polynomial function of the transmit power. By adjusting various parameters, the model encompasses several existing models as special cases, e.g., ideal PA without distortion or the linear PA distortion model in \cite{bjornsonHardware2013} and \cite{korpiFullDuplex2014}. Besides, our model is also applicable to characterize the output signal of the concatenated DPD-PA system. Furthermore, it can be extended to broadband OFDM systems with dissimilar sub-channel power gains.
	
	\item \emph{Characterization of capacity region:} Based on the nonlinear distortion model, we derive the capacity region of a multi-user NOMA system. In a two-user special case, we devise an optimal power allocation algorithm based on Dinkelbach's method to accurately characterize the boundaries of capacity region. Numerical analysis shows that the non-linear PA distortion considerably shrinks the capacity region compared to the ideal case without distortion. In other words, the conventional ideal PA model tends to overestimate the communication performance of NOMA in practice.

    \item \emph{Optimal power control for throughput optimization:} For practical engineering applications, we formulate a general weighted sum rate maximization (WSRMax) problem and propose an efficient fractional programming solution algorithm. For the special case with equal weighting factors, we derive a closed-form expression of the optimal transmit power solution and obtain some interesting insights of the optimal power control policy under PA distortion. Our numerical results show that the optimal power control policy under the practical PA non-linear model offers over 13\% throughput performance gain compared with that under an ideal linear PA assumption.
\end{itemize}
Overall, our results reveal the importance of adapting wireless resource allocation to the nonlinear effect of practical radio frequency (RF) systems, and provide efficient power control methods to attain an optimal throughput performance.

%X. Hong, J. Wang, C. Wang and J. Shi, "Cognitive radio in 5G: a perspective on energy-spectral efficiency trade-off," in IEEE Communications Magazine, vol. 52, no. 7, pp. 46-53, July 2014, doi: 10.1109/MCOM.2014.6852082.
%N. Hassan, S. Gillani, E. Ahmed, I. Yaqoob and M. Imran, "The Role of Edge Computing in Internet of Things," in IEEE Communications Magazine, vol. 56, no. 11, pp. 110-115, November 2018, doi: 10.1109/MCOM.2018.1700906.
%Shaohua Wan, Zonghua Gu, Qiang Ni, Cognitive computing and wireless communications on the edge for healthcare service robots, Computer Communications, Volume 149, 2020, Pages 99-106.
\vspace{-10pt}
\subsection{Related Works}\label{sec2}
1) \textbf{NOMA resource allocation optimization:}
There have been extensive research interests in the performance analysis and resource allocation optimization of NOMA systems \cite{vaeziNOMA2019,coverELEMENTS,tseFundamentals2005,tseMultiaccess1998,weingartenCapacity2006,weiPerformance2020}. The concept of NOMA is essentially an application of SIC decoding process developed for uplink multi-access channels, and superposition coding for downlink broadcast channels\cite{vaeziNOMA2019}. \cite{coverELEMENTS} and \cite{tseFundamentals2005} have derived the capacity regions of Gaussian uplink and downlink NOMA systems. \cite{tseMultiaccess1998} extends the capacity region of the multi-access channel under channel fading, while \cite{weingartenCapacity2006} studies the capacity region of multiple-input multiple-output (MIMO) Gaussian broadcast channels. \cite{weiPerformance2020} compares the performance of NOMA and OMA uplink communication systems for both single-antenna and multi-antenna setups. To maximize the communication capacity, extensive studies have been carried out on resource allocation optimization in NOMA systems. For instance, \cite{timotheouFairness2015} formulates a non-convex problem to maximize downlink throughput under a fairness constraint and develops low-complexity optimal algorithms. \cite{yangOptimality2017} considers maximizing the sum rate performance under individual rate requirement constraints and derives the optimal user decoding order. All the aforementioned works assume an ideal PA in the transmitters and neglect the nonlinear distortion noise. This assumption leads to an overestimation of system throughput performance, and the resulting power allocation method is indeed sub-optimal in practical NOMA systems with PA non-linear distortion as we will show in Section \ref{section5}.

2) \textbf{NOMA under nonlinear PA distortion:} Existing studies have shown that the performance of OFDM systems severely deteriorates in the presence of nonlinear PA distortion \cite{costaImpact1999,dardariTheoretical2000,aggarwalEndtoend2019}. DPD is a widely used method to mitigate the nonlinear distortion effects. However, the use of DPD increases the complexity of transmit RF chain, and it's worth noting that the nonlinearity effect cannot be entirely eliminated at the RF level \cite{qiPower2012}. Therefore, accurate and tractable modeling of the PA non-linear distortion is critical for optimizing the system resource allocation. \cite{liResidual2020} investigates the impact of non-linear PA distortion on cooperative NOMA networks, demonstrating that the asymptotic ergodic capacity becomes saturated as transmit power grows, due to the fact that the PA noise power is related to the transmit power.
%that the distortion noise results an upper bound of achievable capacity.
\cite{guerreiroReceiver2020} considers the impact of nonlinear PA in the NOMA-OFDM system, and proposes a Bussgang receiver to replace the conventional SIC NOMA decoding method to mitigate the non-linear signal distortion. Nonetheless, \cite{liResidual2020} and \cite{guerreiroReceiver2020} assume a simplified model, which assumes that the distortion noise power is proportional to transmit power. The model, although analytically simple, does not accurately capture the non-linear PA effect, especially in the high transmit power region. Besides, none of the existing works are able to model the distortion behavior of PA when DPD is used.

In this paper, we provide accurate and tractable signal modeling of PA distortion based on real-world measurements and propose efficient power control methods for rate maximization in NOMA systems under the considered PA distortion model. The remainder of this paper is organized as follows. In Section \ref{section1}, we present the PA non-linear model and the system model of uplink NOMA under PA distortion. In Section \ref{section2} and \ref{section3}, we characterize the capacity region and propose optimal power control methods that maximize the throughput performance, respectively. We extend the proposed signal model to broadband channels in Section \ref{section4}. In Section VI, we evaluate the performance of the proposed power control algorithms via numerical simulations. Finally, we conclude the paper in Section \ref{section6}.
\vspace{-8pt}
\section{Preliminary and System Model}\label{section1}
In this section, we first briefly introduce the background of non-linear PA distortion and then propose a tractable model based on real-world measurements. Accordingly, we derive the baseband equivalent signal model of uplink NOMA under the proposed PA distortion model, which will be used in subsequent sections for capacity analysis and throughput optimization.
\vspace{-8pt}
\subsection{PA Distortion and DPD}
Consider a concatenated DPD-PA structure at the transmitter of a user equipment (UE). Suppose that a tagged UE transmits a complex baseband data sample sequence $\boldsymbol{u}=(u[1],...,u[n])$. As shown in Fig. \ref{fig:DPD_b}, the data sample sequence is first fed into a DPD module, and the output of the pre-distorted signal is expressed as
\begin{figure}[tbp]
	%	\vspace{-8pt}
	\label{fig:DPD_figure}
	\subfigure[DPD-PA structure.]{
		\centering
		\label{fig:DPD_a}
		\includegraphics[scale=0.6]{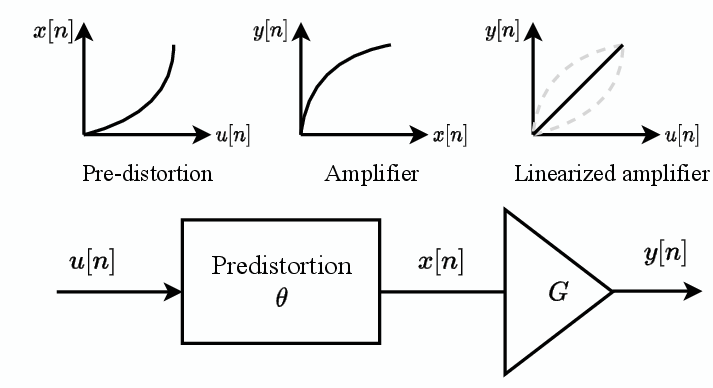}}
	%\\
	\subfigure[Indirected learning of the DPD module.]{
		\centering
		\label{fig:DPD_b}
		\includegraphics[scale=0.6]{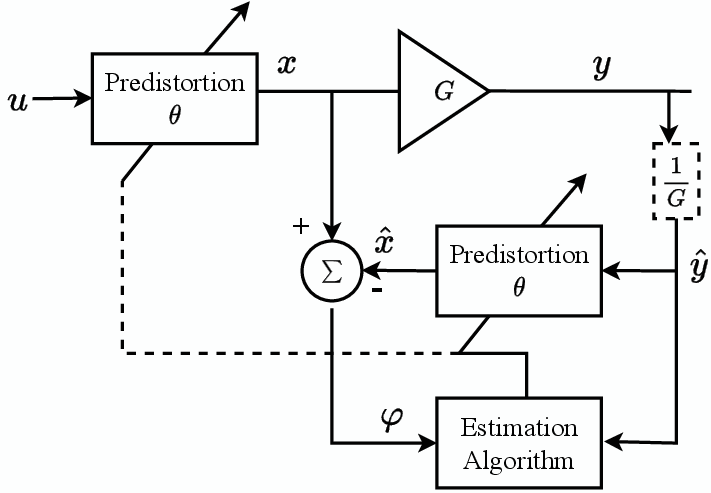} }
	\captionsetup{font=footnotesize}
	\caption{The schematics of DPD and the indirect learning method \cite{chani-cahuanaDigital}.}
	%	\vspace{-1pt}
\end{figure}
\begin{equation}
x[n]=\phi_{\theta} (u[n:n-L]),n=L+1,L+2,\cdots,
\end{equation}
where $\theta$ corresponds to the parameter of the DPD module, and $L$ corresponds to the highest memory depth of the DPD basis functions. In a special case without DPD, we can simply set $x[n] = u[n]$. The parameter $\theta$ is determined through training, utilizing techniques such as direct and indirect training methods. These methods are designed for different PA behavior models, such as memory polynomial (MP) and generalized memory polynomials (GMP) models \cite{chani-cahuanaDigital}.

For a GMP structure, the output of the DPD module is \cite{chani-cahuanaDigital},
%\begin{equation}\label{eq1}
%\begin{split}
%&x[n]=\sum\limits_{k=0}^{K_a-1} \sum\limits_{l=0}^{L_a-1}a_{kl}u[n-l]\lvert u[n-l]\rvert ^k  \\
%&+\sum\limits_{k=1}^{K_b}\sum\limits_{l=0}^{L_b-1}\sum\limits_{m=1}^{M_b}b_{klm} u[n-l]\lvert u[n-l-m]\rvert^k \\
%&+\sum\limits_{k=1}^{K_c}\sum\limits_{l=0}^{L_c-1}\sum\limits_{m=1}^{M_c}c_{klm}u[n-l]\lvert u[n-l+m]\rvert^k,
%\end{split}
%\end{equation}
\begin{equation}\label{eq1}
\begin{split}
&x[n]=\sum\limits_{p=0}^{P_a-1} \sum\limits_{l=0}^{L_a-1}a_{pl}u[n-l]\lvert u[n-l]\rvert ^p  \\
&+\sum\limits_{p=1}^{P_b}\sum\limits_{l=0}^{L_b-1}\sum\limits_{q=1}^{Q_b}b_{plq} u[n-l]\lvert u[n-l-q]\rvert^p \\
&+\sum\limits_{p=1}^{P_c}\sum\limits_{l=0}^{L_c-1}\sum\limits_{q=1}^{Q_c}c_{plq}u[n-l]\lvert u[n-l+q]\rvert^p,
\end{split}
\end{equation}
where $\left\{P_a,P_b,P_c\right\}$ are the highest orders of nonlinearity, $\left\{L_a,L_b,L_c\right\}$ are the highest memory depths, and $\left\{Q_b,Q_c\right\}$ denote the longest lagging and leading delay tap length, respectively. Here, subscripts $a, b,$ and $c$ refer to the aligned signal and envelope, the signal and lagging envelope, and the signal and leading envelope, respectively. $\left\{a_{pl},b_{plq},c_{plq}\right\}$ are the coefficients of the general memory polynomial applied on the aligned terms, the lagging and leading cross-term.
%are the complex coefficients of the aligned signal and envelope, the signal and lagging envelope, and the signal and leading envelope, respectively.
We denote $\boldsymbol{\theta} = (\theta[1],...,\theta[J])^\prime$ as the coefficient vector of DPD by collecting all the coefficients $\left\{a_{pl},b_{plq},c_{plq}\right\}$ into a $J \times 1$ vector. The coefficient vector $\boldsymbol{\theta}$ is initialized, e.g., setting $x[n] = u[n]$, and will be updated iteratively based on the PA output measurements to mitigate the non-linear distortion.

As shown in Fig. \ref{fig:DPD_b}, the pre-distorted signal is then fed into the PA, leading to an output $y[n]$ that is a non-linear transformation of $x[n]$.
After scaling the output as $\hat{y} \!=\! \frac{1}{G} \cdot y $, we use $\hat{y}$ to train the DPD parameters. For the indirect learning method, the estimate of the feedback loop, denoted by $\boldsymbol{\hat{x}} = (\hat{x}(1),...,\hat{x}(N)) $, is
\begin{equation}
\boldsymbol{\hat{x}}=\hat{\boldsymbol{Y}}\boldsymbol{\theta},
\end{equation}
where $\hat{\boldsymbol{Y}} $ is constructed by replacing $u[n]$ in (\ref{eq1}) with $ \hat{y}[n] $, and stacking the $N$ measurements into a $ N \times J $ matrix \cite{chani-cahuanaDigital}. Denote the estimation error as
\begin{equation}
\boldsymbol{\varphi}=\boldsymbol{x}-\boldsymbol{\hat{x}}.
\end{equation}
%Intuitively, we desire $y[n]$ to be equal to $u[n]$, such that by pre-distorting the input sequence $u[n]$ as $\boldsymbol{U}\boldsymbol{w}$ ($\boldsymbol{U} $ is similarly obtained as $ \mathbf{Y} $ by replacing $ y[n]$ with $u[n]$), the PA output is exactly $u[n]$.
%Since the error $e_o$ is linear in the parameters $\boldsymbol{w}$,
The optimal $\boldsymbol{\theta}$ can be obtained by the least squares method,
\begin{equation}
\boldsymbol{\theta}^{\text{opt}}=\arg \ \underset{\boldsymbol{\theta}} \min \lVert \boldsymbol{\varphi} \rVert_2^2,
\end{equation}
and the optimal solution is
\begin{equation}
\boldsymbol{\theta}^{\text{opt}}=\left(\hat{\boldsymbol{Y}}^H \hat{\boldsymbol{Y}}\right)^{-1}\hat{\boldsymbol{Y}}^H \boldsymbol{x}.
\end{equation}
Then, the DPD coefficients are copied from the feedback loop to the DPD module. The iteration repeats until a certain termination condition is met.

The normalized mean square error (NMSE) in dB, defined as follows, is a commonly used metric for characterizing distortion \cite{demirBussgang2021},
\begin{equation}\label{signalmodel}
\eta = 10\lg \left(  \frac{\mathbb{E} \left[\lvert y-Gu \rvert^2\right] }{ \mathbb{E} \left[\lvert y\rvert^2 \right]  }\right) =  10\lg  \left(\frac{\mathbb{E} \left[\lvert e\rvert^2\right]  }{\mathbb{E}\left[ \lvert y\rvert^2\right] }\right),
\end{equation}
where $\mathbb{E}[\cdot]$ is the expectation operation over independent realizations of data samples. Here, we remove the sequence index $n$ for notation brevity. Notice that the PA nonlinear effect is related to the input signal power, where a larger input signal power results in a greater degree of distortion and vice versa. It is necessary to retrain the DPD parameters whenever the signal input power changes significantly. As a result, $\!\eta\!$ varies under different transmit power  $E[|y|^2]$.

\begin{figure}[tbp]
	\centering
	%	\vspace{-8pt}
	\includegraphics[scale=0.6]{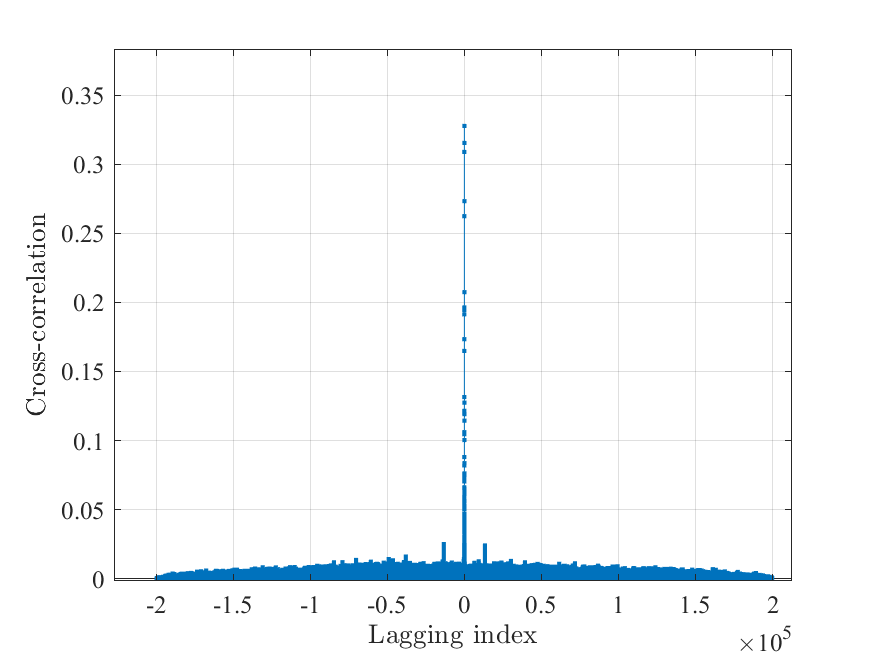}
	\captionsetup{font=footnotesize}
	\caption{The cross-correlation between useful signal and nonlinearity noise signal. The corresponding parameters of the PA distortion model are $(a,\alpha)=(0.0032,1.3552)$, with the operating bandwidth of 30 MHz.}
		\vspace{4pt}
	\label{corr}
\end{figure}

\begin{figure}[tbp]
	\centering
	%	\vspace{-8pt}
	\includegraphics[scale=0.6]{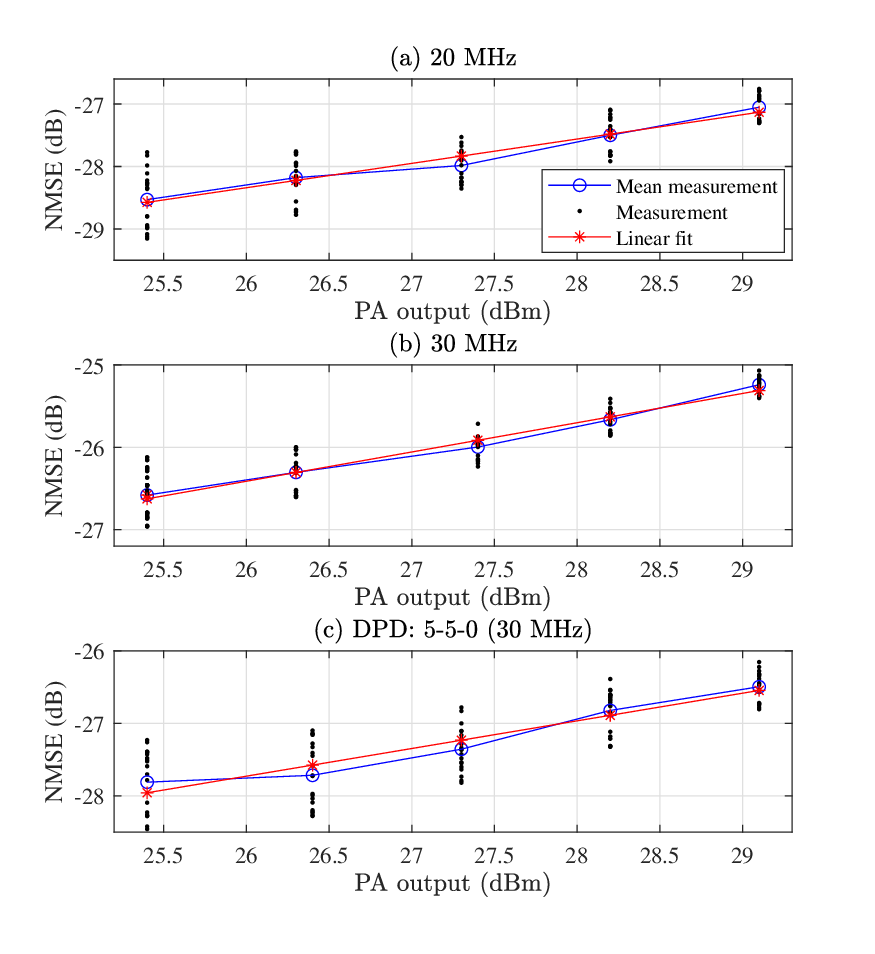}
	\captionsetup{font=footnotesize}
	\caption{ NMSE under different PA output power: (a) the 20 MHz bandwidth case; (b) the 30 MHz bandwidth case and (c) the 30 MHz bandwidth case with 5-5-0 DPD. The corresponding parameters of the PA distortion noise model in (\ref{PN}) are $(a,\alpha)=(0.0021,1.3897)$, $(0.0032,1.3552)$, and $(0.0024,1.3815)$, respectively. The difference of measurement values under the same PA output is caused by the noise of the measurement circuit.}
	\label{NMSE}
\end{figure}
\vspace{-8pt}
\subsection{The Proposed PA Distortion Model}
It is well known that by the Bussgang's Theorem \cite{demirBussgang2021}, when a signal passes by a memoryless distortion function, the output distorted signal can be decomposed as
\begin{equation}\label{bussgang}
y = Gu+e,
\end{equation}
where $e$ denotes the PA distortion noise, and $G$ denotes the Bussgang gain.
Specifically, the useful signal and the non-linear noise are uncorrelated, i.e., $\mathbb{E}$$[u^{*} e]=0$. We plot in Fig. \ref{corr} the correlation of useful signal and distortion. The plot verifies that Bussgang's Theory approximately holds in the memory PA system under consideration. In this case, we have
\begin{equation}
\mathbb{E}\left[ \lvert y\rvert^2\right]
= G^2 \mathbb{E}\left[ \lvert u\rvert^2\right]  + \mathbb{E}\left[ \lvert e\rvert^2\right].
\end{equation}
Using the signal power model in (9), Fig. \ref{NMSE} illustrates the NMSE of a Cree CGH40006 PA\footnote{The details of the PA specifications can be found at http://dpdcompetition.com/rfweblab/} as the output power $P_T $ increases. Specifically, Fig. \ref{NMSE}(a) considers 20 MHz OFDM signal input to the PA without using DPD. Likewise, Fig. \ref{NMSE}(b) and (c) consider scenarios where the signal bandwidth to 30 MHz, input to a PA without DPD and using a 5-5-0 DPD structure ($L_a =  P_a = 5$, no lag/lead term), respectively.
In all three cases, the NMSE appears to increase almost linearly with the output power $P_T$, when plotted on a logarithmic scale, within the given transmit power range.
To better understand these trends, we use linear regression models to fit the measurements. The $R^2$ values of the three regression models are 0.9753, 0.9865, and 0.9510, respectively, indicating high fitting accuracy. Accordingly, we define the following linear function
\begin{equation}\label{TD}
\eta_{TD} = 10\lg \left(\frac{P_N}{P_T}\right)=k_1 \cdot 10\lg (1000 \cdot P_T)+k_2 ,
\end{equation}
with $k_1 \textgreater 0$ and $P_N \triangleq \mathbb{E} \left[ \lvert e\rvert^2 \right]$.
Equivalently, we can express the equation as
\begin{equation}\label{PN}
\begin{split}
P_N= P_T ^{1+k_1} \cdot 1000^{k_1} \cdot 10^{\frac{k_2}{10}} \triangleq a P_{T} ^\alpha.
%\exp \left(\frac{b\ln(10)}{10}\right) \triangleq a\cdot P_T^{\alpha}.
\end{split}
\end{equation}
With a slight abuse of notation and without causing confusions, we denote the output signal of the PA as
\begin{equation}
y = \sqrt{P_T-P_N}\cdot u+e ,
\end{equation}
where $E[|u|^2] =1$ and $e \sim \mathcal{N}(0,P_N)$.
For a practical PA like that in Fig. \ref{NMSE}, $P_N$ is more than two orders of magnitude smaller than $P_T$. Then, the above signal model can be accurately approximated by
\begin{equation}\label{signal_model}
y = \sqrt {P_T} \cdot u +e.
\end{equation}
It is worth noting that the distortion power $P_N$ is not solely dependent on the input power of the PA. The bandwidth also plays a crucial role as a broadband system, with its higher Peak-to-Average Power Ratio (PAPR), tends to generate a greater degree of nonlinear distortion.

\emph{Remark 1}: The model in (\ref{PN}) can encapsulate a range of existing PA models by adjusting the parameters $(\alpha, a)$.
For example, $a =0$ corresponds to an ideal PA without distortion; $\alpha=0$ corresponds to transmit circuit thermal noise of constant power (no PA non-linear distortion); $\alpha=1$ corresponds to the conventional linear PA distortion noise power model \cite{bjornsonHardware2013}; and $\alpha >1$, particularly, corresponds to the practical PA model established based on the measurements in Fig. \ref{NMSE}.

\vspace{-16pt}
\subsection{NOMA Signal Model}
In this subsection, we derive a baseband equivalent signal model for NOMA communication system. We consider a simple uplink NOMA transmission with $K$ UEs transmitting to a base station (BS), where each UE and BS is equipped with a single antenna.
We denote the complex channel coefficient between the BS and UE$_k$ as $h_k$. For the convenience of theoretical analysis, we assume perfect knowledge of channel state information at the BS.\footnote{Pilot sequences can be transmitted with narrow band and low transmit power resulting in negligible PA distortion compared to the transmission of high-rate communication data.}
Besides, we assume that all PAs have identical non-linear behavior. Our work can be readily generalized to the case of heterogeneous PAs by relaxing this assumption.

Considering the signal model in (\ref{signal_model}), let us assume that the baseband signal transmitted by the $k$th UE is $u_k$, and the distortion noise is $e_k$, where $k=1,...,K$. Then, the received baseband equivalent signal at the BS is
%We now focus on the uplink scenario where a set of $M$-users communication to a single receiver in NOMA system. Consider the multiple-access Gaussian channel
\begin{equation}\label{sum signal}
	y_0 = \sum_{k=1}^{K} h_k(\sqrt{p_k} u_k+e_k) +w_0,
\end{equation}
%where $K$ and $w_0$ denote the number of users and the receiver noise with variance $N_0$, respectively.
where $ E[| u_{k} |^2] = 1$ and $p_{k}$ denotes the transmit power. $e_k$ and $u_k$ are uncorrelated zero-mean complex Gaussian random variables with $e_k \sim \mathcal{N}(0,a p_k^{\alpha})$. $w_0 \sim\mathcal{CN}(0, N_0) $ denotes the receiver complex Gaussian noise at the BS. The BS performs successive decoding of the users' information following a predetermined order.
Denote $\pi$ as a decoding order, $\pi(i) $ as the $i$th element of $\pi$. For instance, if $\pi\!  =\!  2\! \rightarrow \! 1\!  \rightarrow\!  3$, then $\pi(1) =2$ and  $\pi(2) =1 $.

Given a specific decoding order $\pi$, the achievable rate of the $k$th decoding user is\cite{tseFundamentals2005}
\begin{equation}\label{rate_vector}\footnotesize
R_{\pi(k)}^{\pi}(\boldsymbol{p})\!\! =\!\!  \log_2\!\!\left(\!\! 1\!\!  +\!\!  \frac{p_{\pi(k)} |h_{\pi(k)}|^2}{\sum\limits_{j=k+1}\limits^{K}p_{\pi(j)}|h_{\pi(j)}|^2 \!+ \!\sum\limits_{i=1}\limits^{K}\! ap_i^{\alpha} |h_i|^2 \!+ \!N_0} \!\!\right)\!\!,  \!\forall k.
\end{equation}
%for $k= 1,\cdots, K$.
Compared with the NOMA model, each user now suffers from an additional interference caused by the PA distortion noise of all users. Meanwhile, the PA distortion term is independent of the decoding order and cannot be eliminated using SIC decoding. This suggests that a throughput analysis of NOMA assuming an ideal PA model might overestimate the real achievable communication rate. In the next section, we will derive the capacity region of NOMA in the presence of non-linear PA distortion.
\vspace{-16pt}
\section{Characterization of Capacity Region with Nonlinear PA Distortion Power}\label{section2}
In this section, for clarity and ease of explanation, we first derive the capacity region of a two-user NOMA system with PA distortion power. Subsequently, we extend this result to accommodate a general multi-user system where $K > 2$.
\vspace{-20pt}
\subsection{Capacity Analysis of Two-user Case}
\vspace{-5pt}
When $K=2$, the received baseband equivalent signal at the BS is
\begin{equation}\label{system model}
y_{1} = h_1(\sqrt{p_1} u_1+e_1)+h_2(\sqrt{p_2} u_2+e_2)+w_0.
\end{equation}
The BS uses SIC to decode the messages of the two UEs sequentially. We first consider a decoding order $\pi_1 = 2\! \rightarrow \! 1$. From \eqref{rate_vector}, the achievable rate of the two UEs are
\begin{equation}\label{1}
\begin{split}
&R_1^{\pi_1} = \log_2\left(1+\frac{p_1|h_{1}|^{2} }{a p_1^\alpha |h_{1}|^{2} + a p_2^\alpha |h_{2}|^{2} + N_0}\right),\\
&R_2^{\pi_1} = \log_2\left(1+\frac{p_2|h_{2}|^{2} }{ (p_1+ a p_1^\alpha) |h_{1}|^{2} + a p_2^\alpha |h_{2}|^{2} + N_0} \right).
\end{split}
\end{equation}
Similarly, the achievable rates of the two UEs under the other decoding order $\pi_2$ = 1 $\rightarrow$ 2 are
\begin{equation}\label{2}
\begin{split}
&R_1^{\pi_2} = \log_2\left(1+\frac{p_1|h_{1}|^{2} }{  a p_1^\alpha |h_{1}|^{2} + (p_2 + a p_2^\alpha)|h_{2}|^{2} + N_0}\right),\\
&R_2^{\pi_2} =\log_2\left(1+\frac{p_2|h_{2}|^{2} }{a p_1^\alpha |h_{1}|^{2} + a p_2^\alpha |h_{2}|^{2} + N_0}\right).
\end{split}
\end{equation}
%According to (\ref{1}), the receiver decodes the information of $UE_{2}$ firstly while treating the signal from $UE_{1}$ as Gaussian interference, then the receiver reconstruct $UE_{2}$'s information and remove it from the total receive signal. Once signal from $UE_{2}$ can be decoded successfully, $UE_{1}$ is able to transmit at the rate $R_1^{\pi_1}$.
In equations (\ref{1}) and (\ref{2}), one can note that the distortion noises $e_1$ and $e_2$ cannot be eliminated by SIC, which means each user experiences the PA distortion noise from both users.
%Once signal from $UE_{1}$ can be decoded successfully, $UE_{2}$ thus can decoding its information while exiting only thermal noise and vice versa in (\ref{2}).
%The denominator term means the noise in receiver consist of not only the additive thermal noise but also the nonlinear PA distortion from each independent antenna.
%=====================total rate===========
Meanwhile, the sum throughput is given by
\begin{equation}\label{s1}
\begin{split}
\begin{aligned}
R_1^{\pi_1}+R_2^{\pi_1}&=R_1^{\pi_2}+R_2^{\pi_2}\\
&=\log_2 \left(1+ \frac{p_1|h_{1}|^{2}+p_2|h_{2}|^{2}}{ a p_1^\alpha |h_{1}|^{2} + a p_2^\alpha|h_{2}|^{2}\textbf{} + N_0} \right).&
\end{aligned}
\end{split}
\end{equation}
We can see in (\ref{s1}) that the sum rate is irrelevant to the decoding order similar to the conventional NOMA scheme that does not consider PA distortion noise.

From (\ref{1}) and (\ref{2}), the achievable rate region of the uplink NOMA under the decoding order $\pi_q$, denoted by $\mathcal{R}^{\pi_q}$, is
% \cite{satoTwouser1977,rMultiway1971}
\begin{equation}\label{achievable}
\mathcal{R}^{\pi_q} =: \bigcup\limits_{p_1,p_2} \left\{ (R_1^{\pi_{q}}, R_2^{\pi_{q}})|p_1,p_2 \leq \bar{p} \right\}, q = 1,2,
\end{equation}
where $\bar{p}$ denotes the maximum transmit power, which is assumed to be equal for the two UEs.
In this section, we characterize the achievable rate pairs in the boundary of $\mathcal{R}^{\pi_q}$, where each UE cannot further improve its rate without degrading the other's, pointing to the optimal rate trade-off. Without loss of generality, we consider the decoding order $\pi_{1}$ in the following analysis.

Consider an extreme case where only one UE transmits while the other remains silent. Let $R_k^{\rm{max}}$ denote the maximum transmission rate of the point-to-point link for UE$_k$. We derive its closed-form expression in the following Theorem 1.

\textbf{Theorem 1:} When $\alpha \in [0,1]$, the maximum point-to-point rate of UE$_k$ is
\begin{equation}\label{opt1}
R_k^{{\rm max}}=\log_2\left(1+\frac{\bar{p}|h_{k}|^{2} }{a \bar{p}^\alpha |h_{k}|^{2} + N_0}\right), k = 1,2.
\end{equation}
When $\alpha >1$, the maximum rate is
%\begin{equation}
\begin{small} \footnotesize
	\begin{subnumcases}
	%\begin{split}
	{R_k^{{\rm max}}\!\!\!=\!\!}
	\!\!\!\log_2\!\!\left(\!\!1 \!\! + \!\!\frac{\bar{p_k}|h_{k}|^{2} }{a \bar{p_k}^\alpha |h_{k}|^{2} + N_0}\right), \!\!\!\!\!\!&\!\!if \  $p_k^{\text{opt}} > \bar{p}$,  \\
	\!\!\!\log_2\!\! \left( \!\!1\!\!+ \!\! \frac{{\frac{N_0}{\alpha}} |h_k|^2 }{{\frac{N_0}{\alpha}} a |h_k|^2 + N_0  \left( a(\alpha-1)|h_k|^2 \right) ^{\frac{1}{\alpha} } } \right)\!\!, \!\!& otherwise,\label{eq4}
	\end{subnumcases}
\end{small}
%\end{equation}
where $p_k^{\text{opt}}= \sqrt[\alpha]{ \frac{N_0}{a\left(\alpha-1\right)|h_{k}|^{2}}}$.

\emph{proof}:
	The achievable rate of the point-to-point uplink link is
	\begin{equation}\label{Ri}
	R_k(p_k) = \log_2\left(1+ \gamma(p_k) \right),
	\end{equation}
	with $\gamma \left( p_k\right) =\frac{p_k|h_k|^{2} }{a p_k^\alpha |h_k|^{2} + N_0}$. We then divide our analysis into the following cases.
	%	and $p$ denotes the transmit power, $h$ is the channel coefficient between the user and BS.
	\begin{itemize}
		\item $\alpha=0$.
		%		The distortion term can be seen as transmit circuit thermal noise and
		We have $\gamma \left( p_k\right) =\frac{p_k|h_k|^{2} }{a \cdot |h_k|^{2} + N_0}$. Here, by setting the power allocation as $p_k=\bar{p}$, we obtain the maximum achievable rate $ R_k^{\rm{max}}= \log_2\left(1+\frac{\bar{p}|h_{k}|^{2} }{a |h_{k}|^{2} + N_0}\right). $
		
		%		\boldsymbol{\gamma}
		\item $0 \textless \alpha \textless 1$. $\gamma \left( p_k\right)$ is monotonically increasing as $\gamma'\left(p_k\right) \geq 0$, where $\gamma^\prime = \frac{\partial \gamma}{\partial p_k}$ is the first derivative of $\gamma$ $p_k$. In this case, the maximum achievable rate  $ R_k^{\rm{max}}= \log_2\left(1+\frac{\bar{p}|h_{k}|^{2} }{a \bar{p}^{\alpha}|h_{k}|^{2} + N_0}\right) \triangleq \bar{R}_k $ is achieved by setting $p_k = \bar{p}$.
		
		\item $\alpha=1$.
		%		This corresponds to conventional linear distortion noise power.
		$\gamma\left( p_k\right)$ increases monotonically with $p_k$. Therefore, the maximum achievable rate is also $ R_k^{\rm{max}}= \bar{R_k}$, with optimal transmit power $p_k = \bar{p}$. 	
		\item $\alpha \ \textgreater \ 1$.
		%		This corresponds to the practical PA model in Fig. 2.
		Taking the derivative of $\gamma$ with $p_k$, we get
				\begin{equation}\label{26}
			\gamma^\prime(p_k) = \frac{\partial \gamma}{\partial p_k}=\frac{(ap_{k}^{\alpha}+ N_0)|h_{k}|^2-ap_{k}^{\alpha}\alpha |h_{k}|^4}{(ap_{k}^{\alpha} |h_{k}|^2+N_0)}.
				\end{equation}
		By setting $\gamma^\prime(p_k)$ to zero, we obtain the unique stationary point \begin{equation}\label{27}
		p_{k}^{\text{opt}}=\sqrt[\alpha]{ \frac{N_0}{a\left(\alpha-1\right)|h_{k}|^{2}}}.
		\end{equation}
		\eqref{26} and \eqref{27} demonstrate that when $p_k \leq p_k^{\text{opt}}$,  $\gamma^\prime(p_k) \geq 0$ and $\gamma(p_k)$ increases with $p_k$. Otherwise, when $p_k \textgreater p_k^{\text{opt}}$,  $\gamma^\prime(p_k) \textless 0$ and $\gamma(p_k)$ decreases with $p_k$.
%		From $\gamma^\prime(p_k)$, when $p_{k} \leq p_{k}^{\text{opt}} $, $\gamma\left( p_{k}\right)$ increases with $p_k$. Otherwise, when $p_{k} > p_{k}^{\text{opt}} $, $\gamma\left( p_{k}\right)$ decreases with $p_k$.
		Therefore, $p_{k} = p_{k}^{\text{opt}}$ achieves the global maximum of function $\gamma\left( p_k \right)$.
		%		Notice that the global maximum points $p^{*}_i$ are not necessary achievable since the maximum transmit power of UE is $\bar{p_i}$. Therefore, we separate our analysis to two case: $p^{*}_i>\bar{p}$ and $p^{*}_i \leq \bar{p}$. When
Consequently,  $p^{\text{opt}}_k$ is larger than the maximum transmit $\bar{p}$, the maximum rate $ R_k^{\rm{max}}= \bar{R}_k$ is achieved by setting $p_k = \bar{p}$. Otherwise, when $p^{\text{opt}}_k \leq \bar{p}$, the maximum rate in \eqref{eq4} is achieved by setting $p_k=p^{\text{opt}}$.
	\end{itemize}	
This proves the theorem.
		\begin{figure}[tbp]
			\centering
			%	\vspace{-8pt}
			\includegraphics[scale=0.6]{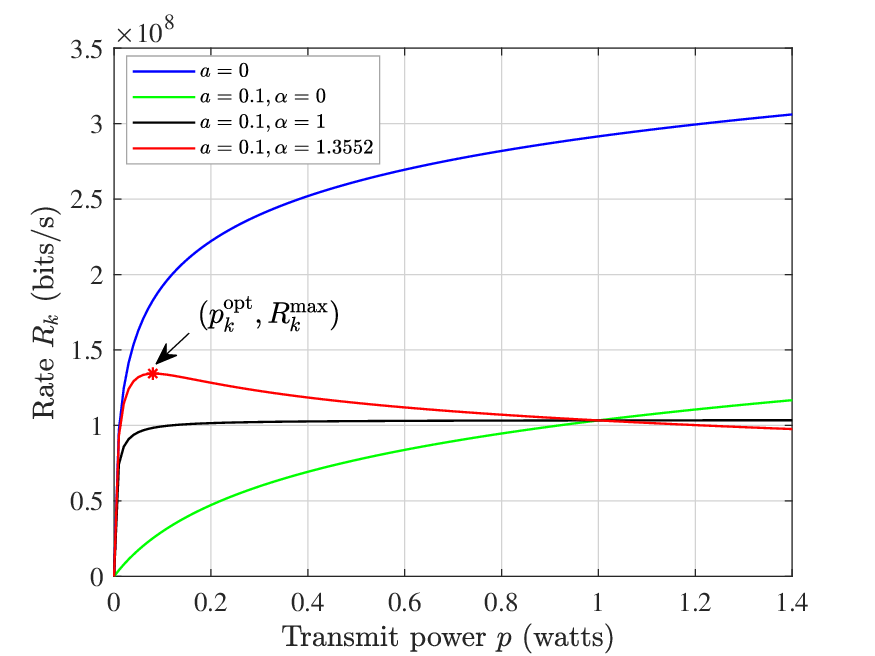}
			\captionsetup{font=footnotesize}
			\caption{An illustration of transmission rate curves under different distortion model parameters $(a,\alpha)$ in \eqref{PN}.}
			%	\vspace{-1pt}
			\label{a}
		\end{figure}
$\hfill \qedsymbol$

	In Fig. \ref{a}, we illustrate the variation of data rate with respect to the transmit power under different values of $(a, \alpha)$. Specifically, $a = 0$ corresponds to the distortion-free model. $(a =0.1 ,\alpha = 0 )$ corresponds to the transmitter thermal noise. $(a = 0.1, \alpha =1)$ corresponds to a linear noise power model. We observe that the data rate increases monotonically  with transmit power except for the curve representing the polynomial power PA distortion model.

\emph{Remark 2}: In the presence of the polynomial PA distortion noise power (i.e., when $\alpha >1$), the maximum rate is in general not attained at the maximum transmit power $\bar{p}$, unlike the case without distortion ($\alpha =0$) or with linear distortion noise power ($\alpha=1$). Instead, the transmitter may reduce the power to alleviate the self-interference caused by PA distortion. In particular, the optimal power $p_k^{\text{opt}}$ decreases with $|h_{k}|^2$, indicating that self-interference dominates the SNR when the link quality is good. We can also infer that the PA will not transmit at a power level larger than $p_k^{\text{opt}}$, as it not only consumes extra energy but also generates higher detrimental interference to the other co-channel NOMA user. Consequently, we can safely conclude that the data rate also increases with transmit power for $\alpha>1$ by confining $p_k \leq p_k^{\text{opt}}$. Notice that the practical PA model established from the measurements in Fig. \ref{NMSE} corresponds to the case of $\alpha>1$. Therefore, we shall only consider computing the achievable rate with $\alpha>1$ in the the subsequent analysis.

We can now characterize the rate pair in \eqref{1} under decoding order $\pi_1$ following the \emph{rate profile} method. That is, we consider
%we assume that $UE_2$ transmit at a rate of $R_2^{\text{max}}$ restricted by information theory, denoted by $\tau$ as a \emph{Rate-Profile}. Then we decrease $\tau$ from $R_2^{\text{max}}$ to zero and solving the following optimization problem:
\begin{subequations}\label{P1}
	\begin{align}
	\text{(P1)}:\ R_{1}^{\pi_1}\left( \tau \right)=\ \ \ & \underset{p_1,p_2}{\text{maximize}}\ \ \ \ \ \ \ \  R_1^{\pi_1} \\
	& \text{subject to}\ \ \ \ \ \ \ \  R_2^{\pi_1} \geq \tau, \\
	%	 \log_2\left(1+\frac{p_2|h_{2}|^{2} }{(p_1+a p_1^\alpha) |h_{1}|^{2} + a p_2^\alpha |h_{2}|^{2} + \sigma_0^2}\right) \geq \tau\\
	& \ \ \ \ \ \ \ \ \ \ \ \ \ \ \ \ \ \ \ \  0\leq p_1, p_2 \leq \bar{p},
	\end{align}
\end{subequations}
where $\tau \in \left[ 0,R_2^{\text{max}} \right]$.
From Remark 2, we can safely replace the maximum power constraint (\ref{P1}c) by $p_k \leq \min\{\bar{p}, p_k^{\text{opt}}\}$ without affecting the optimal solution of (P1), such that $R_k^{\pi_1}$ increases with $p_k$ under the new power constraint.  This also indicates that the equality in (\ref{P1}b) must hold at the optimum, because otherwise, we can reduce the transmit power $p_2$ and attain a higher objective.
%Accordingly, for each $\tau$, we obtain a boundary point $\left(\tau,R_1^{\pi_1}(\tau) \right)$ by solving (P1).
Accordingly, for each $\tau$, we obtain a point $\left(\tau,R_1^{\pi_1}(\tau) \right)$ on the boundary of $\mathcal{R}^{\pi_{1}}$ by solving (P1).
Then, by varying  $\tau$ from 0 to $R_2^{\text{max}}$, the entire boundary of the achievable rate region can be found by solving (P1) repeatedly. Note that (P1) is a non-convex problem because of the fractional terms in the objective. In the following, we apply Dinkelbach's method to solve (P1).
 \allowdisplaybreaks
We define the received SINR of $R_1^{\pi_{1}}$ as
\begin{equation}\label{y1}
\gamma_1=\frac{p_1|h_{1}|^{2}}{a p_1^\alpha |h_{1}|^{2} + a p_2^\alpha |h_{2}|^{2} + N_0},
\end{equation}
%\begin{equation}\label{y2}
%\gamma_2=\frac{p_2|h_{2}|^{2}}{a p_1^\alpha |h_{1}|^{2} + (p_2 +a p_2^\alpha) |h_{2}|^{2} + \sigma_0^2}
%\end{equation}
and $R_1^{\pi_{1}}=\log_2\left(1+\gamma_1 \right)$. Then, (P1) is equivalent to maximizing $\gamma_1$ with constraint (\ref{P1}b) rewritten as
\begin{equation}\label{constrain1}
p_2|h_2|^2- \left(2^{\tau} -1\right) \left( (a p_1^\alpha +p_1)|h_{1}|^{2} + a p_2^\alpha |h_{2}|^{2} + N_0\right)   \geq 0,
\end{equation}
\eqref{constrain1} is a convex constraint with respect to $p_1,p_2$ under a constant $\tau$ when $\alpha>1$. Accordingly, (P1) is equivalently rewritten as
\begin{subequations}\label{P2}
	\begin{align}
	\ \ \ & \underset{p_1,p_2}{\text{maximize}}\ \ \ \ \ \ \ \  \gamma_1 \\
	& \text{subject to}\ \ \ \ \ \ \ \  \text{(\ref{P1}c), (\ref{constrain1})}.
	\end{align}
\end{subequations}
Although the objective function $\gamma_1$ is non-convex, its numerator is concave and the denominator is convex. Such a problem can be efficiently tackled by the Dinkelbach's method for fractional programming \cite{shenFractional2018}. By applying the Dinkelbach's transform to the objective, (\ref{P2}) can be equivalently written as
\begin{subequations}\label{P5}
	\begin{align}
	R_{1}^{\pi_1}\left( \tau \right)= \  & \underset{p_1,p_2}{\text{maximize}}\ \ \ \  F\\
	& \text{subject to}\ \  \text{(\ref{P1}c), (\ref{constrain1})},
	\end{align}
\end{subequations}
where
\begin{equation}\label{F}
F = p_1|h_1|^2- \gamma_1 \left( a p_1^\alpha |h_{1}|^{2} + a p_2^\alpha |h_{2}|^{2} + N_0 \right).
\end{equation}
$\gamma_1$ is an auxiliary variable, which is exactly in the form of (\ref{y1}). Given $\gamma_1$, (\ref{P5}) is a standard convex optimization problem that can be efficiently solved. The optimal solution is then used to update $\gamma_1$. We summarize the details of solving (P1) in Algorithm 1. In each iteration, we solve (\ref{P5}) and then update $\gamma_1$ and $F$ according to (\ref{y1}) and (\ref{F}), until a convergence condition is met. The boundary of the achievable pairs of $\mathcal{R}^{\pi_1}$ can be obtained by traversing $\tau \in [0,R_2^{\text{max}}]$. Subsequently by reversing the decoding order, we can obtain $\mathcal{R}^{\pi_2}$ using a similar method.
The Dinkelbach's method guarantees convergence to a global optimum solution for a single-ratio fractional programming problem such as (P1) \cite{shenFractional2018}. So, the proposed characterization of achievable rate boundaries is exact.
%
%%Problem(\ref{P5}) is a standard convex optimization problem and thus can be solved by CVX tool.
%As a result, the optimization problem for characte]rize achievable rate region is solved as a standard convex optimization problem in an iterative manner with double loops. In the outer loop, we traverse $r_2$ in the range of $ [0, R_2^{\text{max}}] $. In the inner loop, we solve  (\ref{P5}) with the Dinkelbach's algorithm, updating $\gamma_1$ according to (\ref{y1}), until the gap between zero and value of objective function is below a small threshold $\epsilon$. Then we reverse the decoding order and obtain $\mathcal{R}^{\pi_2}$. The details of the proposed algorithm are presented in the following algorithm.

\begin{algorithm}
	%	\small
	%		\scriptsize
	\SetAlgoLined
	\SetKwData{Left}{left}\SetKwData{This}{this}\SetKwData{Up}{up}
	\SetKwRepeat{doWhile}{do}{while}
	\SetKwFunction{Union}{Union}\SetKwFunction{FindCompress}{FindCompress}
	\SetKwInOut{Input}{input}\SetKwInOut{Output}{output}
	
	\textbf{initialization:} initialize $(p_1,p_2)$ to a feasible value, precision $\epsilon \leftarrow 10^{-4}$;\\
	
	\Repeat{$|F|\leq \epsilon$ } {
		Update the auxiliary variable $\gamma_1$ by (\ref{y1});\\
		Update the objective value $F$ by (\ref{F});\\
		Update $(p_1^{*}, p_2^{*})$ by solving (\ref{P5});\\
	}
	\textbf{return} optimal solution $(p_1^{*}, p_2^{*} ).$
	\caption{Dinkelbach' algorithm for solving (P1)}
\end{algorithm}
	\vspace{-8pt}
%Consequently, the upper bound $R_2^{\text{max}}$ of $r_2$, which indicates the maximum point-to-point communication rate, is given by proposition 1.
\vspace{-16pt}
\subsection{Capacity Region with Two-user NOMA}
We move on to characterize the capacity region of the two-user uplink NOMA system with nonlinear PA distortion power based on the achievable rate boundaries obtained in the last subsection.
To highlight the effect of the non-linear PA distortion, we consider the following three schemes for comparison: 1) NOMA with nonlinear distortion power model (NDM): $P_N = a P_T ^{\alpha}$ in (\ref{PN}), $\alpha>1$;
2) NOMA with linear distortion power model (LDP) in \cite{bjornsonHardware2013}: $\alpha = 1 $ in \eqref{PN};
3) NOMA with an ideal PA (IDEAL): $P_N =0$, i.e., the well-studied uplink NOMA system without PA distortion.
%For convenience, we denote the optimal power control policy under NDM and ideal PA as PC-NDM and PC-IDEAL, respectively.
\begin{figure}[tbp]
	\centering
	\vspace{2pt}
	\includegraphics[scale=0.6]{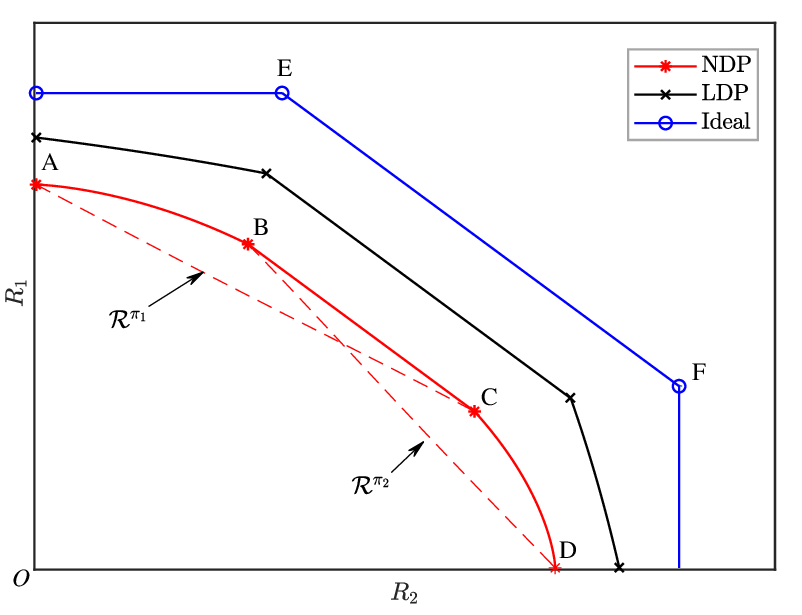}
	\captionsetup{font=footnotesize}
	\caption{Capacity regions with difference distortion power model, where points B, C and E, F correspond to the corner points of NDM and IDEAL capacity regions, respectively.}
	\label{L1}
\end{figure}

As shown in Fig. \ref{L1}, the curve A-B-D denotes the boundary of $\mathcal{R}^{\pi_{1}}$ and the curve A-C-D denotes the boundary of $\mathcal{R}^{\pi_{2}}$ obtained from Algorithm 1. Notice that the boundaries A-B-D and A-C-D are not necessarily convex curves. We can compute capacity region of NOMA-NDM by computing a convex hull of the two boundaries. Specifically, the capacity region, denoted by $\mathcal{C}^N$, is
\begin{equation}\label{convexhull}
\mathcal{C}^N  =: {\rm convex \ hull \ of\  }\left( \mathcal{R}^{\pi_1}\bigcup	\mathcal{R}^{\pi_2}\right),
\end{equation}
and illustrated in Fig. 3 as O-A-B-C-D-O, where the points in the segment B-C are achievable through time-sharing between two operating points under different decoding orders.
In the following, we then show that the segment B-C can be characterized in closed-form by solving the sum rate maximization problem.

\textbf{Proposition 1:} Considering the sum rate maximization problem under decoding order $\pi_1$:
\begin{subequations}\label{sum rate_problem}
	\begin{align}
	b^* =\ \ \ & \underset{p_1,p_2}{\text{maximize}}\ \ \ \ \ \ R_1^{\pi_{1}}+R_2^{\pi_{1}}\\
	& \text{subject to}\ \ \ \ \ \ 0\leq p_1,p_2 \leq \hat{p}_k,
	\end{align}
\end{subequations}
where $\hat{p}_k \!\!\!=\!\!\! \min\left[\bar{p},p^{\text{opt}}_k\right]$.
We define a line $l_{sum}^* \!=\!: \left\{(R_1,R_2)| R_1+R_2 = b^* \right\}$. Then, $l_{sum}^*$ is tangent to regions $\mathcal{R}^{\pi_{1}}$ and $\mathcal{R}^{\pi_{2}}$, forming a segment of the capacity region defined in (\ref{convexhull}).

\emph{Proof}:			
		Please refer to Appendix \ref{app1} for details.
$\hfill \qedsymbol$
%and their intersection can be achieved by time-sharing between the multiple access strategies.

\emph{Remark 3: }
%怎么优化（31）需要先简单说明一下。
%By using an algorithm similar to Algorithm 1 with a Dinkelbach's method, problem (\ref{sum rate_problem}) can be solved by equivalently optimizing its SINR, since
The SINR of objective function of (\ref{sum rate_problem}), as given in (\ref{s1}), is a concave-convex fractional function. Therefore, (\ref{sum rate_problem}) can be exactly and efficiently solved using the Dinkelbach's method, with the details outlined in Section \ref{sec4.a}. From Proposition 1, we infer that the corner point B can be obtained by solving (\ref{sum rate_problem}), and point C is obtained by substituting the decoding order in (\ref{sum rate_problem}) by $\pi_2$. All the other rate points on the segment B-C can be obtained by time-sharing between the two NOMA strategies of points B and C.

Fig. \ref{L1} illustrates the capacity regions of the three considered schemes. With an IDEAL PA, the capacity region is known to be a pentagon \cite{tseFundamentals2005}, where one user can increase its transmission rate from zero to reach a corner point (E or F), while the other user remains at a maximum rate.
For the practical non-linear noise power model under consideration, the capacity region further contracts due to the increased distortion, and the boundary is no longer an intersection of straight lines as in the IDAEL PA case. Notice that the capacity region of the NDM case represents the actual achievable rate pairs of practical PAs. In other words, conventional IDEAL PA indeed overestimates the communication performance of NOMA.
\allowdisplaybreaks
\vspace{-8pt}
\subsection{Extension to Multi-user Case}
The capacity region analysis can be directly extended to multi-user cases.
The boundary of achievable rate region under a decoding order $\pi$ is obtained by
considering the following problem with a given $\tau_k \in [0, R_k^{\text{max}}],  k = 2,3,...,K $,
\begin{subequations}\label{P3}
	\begin{align}
		\text{(P2)}:\ R_{1}^{\pi}\left( \tau \right)=\  & \underset{\boldsymbol{p}}{\text{maximize}}\ \ \ \ R_1^{\pi} \\
		& \text{subject to}\ \ \  R_k^{\pi} \geq \tau_k,  \\
		& \ \ \ \ \ \ \ \ \ \ \ \ \ \ \ 0\leq p_k \leq \bar{p}, \ k = 2,3,\cdots ,K,
	\end{align}
\end{subequations}
where $R_k^{\text{max}}$ given in Theorem 1 denotes the maximum point-to-point link rate and $\boldsymbol{p} = (p_1,...p_k)$. If the above problem is feasible, equality must hold for all the constraints in (\ref{P3}b) at the optimum following similar arguments in (P1).
Then, we can obtain the rate pair $(R_1^{\pi *},\tau_2,...,\tau_K)$ on the boundary of $\mathcal{\pi}$ by using the Dinkelbach's algorithm, where $R_1^{\pi *}$ is the optimal solution of (P2);
%Then, denote $R_1^{\pi*}$ as  (P2), which can be obtained using the Dinkelbach's algorithm, the rate-tuple  $(R_1^{\pi*}, \tau_2,...,\tau_K)$ is on the boundary of $\mathcal{R}^\pi$;
otherwise, we set $R_k^{{\pi}*} = -\infty$ for an infeasible problem. By traversing all feasible $(\tau_2,\cdots,\tau_K)$, we can characterize the boundary of the achievable rate region under the decoding order $\pi$, denoted as
%\begin{algorithm}
%	\SetAlgoLined
%	\SetKwData{Left}{left}\SetKwData{This}{this}\SetKwData{Up}{up}
%	\SetKwRepeat{doWhile}{do}{while}
%	\SetKwFunction{Union}{Union}\SetKwFunction{FindCompress}{FindCompress}
%	\SetKwInOut{Input}{input}\SetKwInOut{Output}{output}
%	
%	\textbf{initialization:} initialize $\tau_2 =...=  \tau_k = 0,\forall k \in \left\{2,...,K \right\} $; \\
%    {
%		\Repeat{\rm $R_1^{{\pi}*} = -\infty$  or $\tau_k \geq R_k^{\rm{max}},\forall k \in \left\{2,...,K \right\}$ } {
%			Solve (P2) by Dinkelbach's algorithlm;\\
%			Update $\tau_k = \tau_k +\delta  , \forall k \in \left\{2,...,K \right\}  $, \rm where $\delta$ is a small positive constant to control the algorithm accuracy;\\}
%	}
%	\textbf{return} optimal solution $\boldsymbol{p}^*$.
%	\caption{Exhaustive search for obtaining $\mathcal{R}^\pi$}
%\end{algorithm}
\begin{equation}\label{achievable_multi}
\mathcal{R}^{\pi}(\boldsymbol{p}) =: \bigcup\limits_{\boldsymbol{p}} \left\{ (R_1^{\pi}, ..., R_k^{\pi})|p_k \leq \bar{p} \right\}, \forall k \in \left\{ 1,...,K \right\} .
\end{equation}
Therefore, the capacity region of the multi-user NOMA system is
\begin{equation}
\mathcal{C} =: {\rm convex \ hull \ of\  }\left(\bigcup\limits_{\pi \in \Pi}	\mathcal{R}^{\pi}(\boldsymbol{p})\right),
\end{equation}
where $\Pi$ is the collection of all possible permutations of the ordered set $\left\{1,...,K \right\}$.
\vspace{-10pt}
\section{Throughput Maximization with Nonlinear PA Distortion Power}\label{section3}

\subsection{Weighted Sum Rate Maximization}
To ensure fair and efficient network operation, it is crucial for a practical communication system to select a suitable operating point from the achievable capacity region. This is typically accomplished by solving a weighted sum rate maximization (WSRMax) problem \cite{weeraddanaWeighted2011}. In this section, we study the following WSRMax problem of the NOMA system with nonlinear PA distortion,
\begin{subequations}\label{P8}
	\begin{align}
		\text{(P3)}:\ \ \ & \underset{\boldsymbol{p}}{\text{maximize}}\ \ \ \ \ \ \ F_o(\boldsymbol{p}) = \sum_{k=1}^{K} \omega_k R_k \\
		& \text{subject to}\ \ \ \ \ \ R_k \geq r_k,  \forall k,\\
        & \ \ \ \ \ \ \ \ \ \  \ \ \ \ \ \ \ \ 0\leq p_k\leq\hat{p_k},
	\end{align}
\end{subequations}
%where $R_k$ is shown in (\ref{rate_vector}).
where $\{\omega_k \textgreater 0,\forall k|\sum_{k=1}^{K}\omega_k=1\}$ are the weighting factors, $ \hat{p_k} = \min\{\bar{p},p_k^{\text{opt}} \} $.
For analytical convenience, we consider the decoding order $\pi(i)=i$ such that the notation $\pi(i)$ in \eqref{rate_vector} can be replaced by $i$.
(\ref{P8}b) represents that $R_k$ should be larger than a prescribed threshold $r_k$.
(P3) is an NP-hard non-convex optimization problem \cite{luoDynamic2008}.
To deal with (P3), we introduce auxiliary variables $\boldsymbol{\gamma} = \{\gamma_k, \forall k\}$, where
\begin{equation}\label{gammak}
\gamma_k = \frac{p_k |h_k|^2}{\sum_{i=k+1}^{K}p_i |h_i|^2 + \sum_{i=1}^{K} ap_i^{\alpha} |h_i|^2 + N_0}.
\end{equation}
\vspace{6pt}
By leveraging the Lagrangian dual transform \cite{khanOptimizing2020}, we can equivalently rewrite (P3) as
\begin{subequations}\label{P9}
	\begin{align}
%	\begin{split}
	 & \underset{\boldsymbol{p}, \boldsymbol{\gamma}}{\text{maximize}}\ \ \  F_l(\boldsymbol{p},\boldsymbol{\gamma} )\\
& \text{subject to} \  \ \ G_k(\boldsymbol{p})  \geq 0, \forall k, \\
& \ \ \ \ \ \ \ \ \ \ \ \ \ \ 0\leq p_k\leq\hat{p_k},\forall k,
%	\end{split}
	\end{align}
\end{subequations}
where
\begin{equation}\label{fr}
\begin{split}
F_l(\boldsymbol{p}, \boldsymbol{\gamma} )&=  \sum_{k=1}^{K} \omega_k \log_2\left(1+ \gamma_k \right)
-\sum_{k=1}^{K} \omega_k \gamma_k\\ & +  \sum_{k=1}^{K} \frac{\omega_k (\gamma_k+1)|h_k|^2 p_k}{\sum_{i=k}^{K} |h_i|^2 p_i + \sum_{i=1}^{K} ap_i^{\alpha} |h_k|^2  +N_0},
\end{split}
\end{equation}
and
\begin{equation}\footnotesize
G_k(\boldsymbol{p}) = p_k|h_k|^2\!\! - \!\!\left(2^{r_k}\!\! -1 \!\right)\!\left( \sum_{i=k+1}^{K} p_i |h_i|^2 + \sum_{i=1}^{K} ap_i^{\alpha} |h_i|^2+N_0 \right).
\end{equation}
Problem (\ref{P9}) is still non-convex due to the summation of fractional terms in (\ref{fr}). Nevertheless, we can solve it using the quadratic transform technique \cite{shenFractional2018a}. To start with, we introduce auxiliary variables $\boldsymbol{y}=\{ y_k\textgreater0,\forall k\} $ and transform $F_l(\boldsymbol{p}, \boldsymbol{\gamma} )$ as
\vspace{-8pt}
\begin{equation}\label{fq}
\begin{aligned}
F_q(\boldsymbol{p}, \boldsymbol{\gamma}, \boldsymbol{y})&  = \sum_{k=1}^{K} \omega_k \log_2(1+ \gamma_k ) -\sum_{k=1}^{K} \omega_k \gamma_k \\
&	+ \sum_{k=1}^{K} \bigg( 2 y_k \sqrt{\omega_k (\gamma_k+1)|h_k|^2 p_k} - \\
&  y_k^2 \Big(	\sum_{i=k}^{K} p_i |h_i|^2 + \sum_{i=1}^{K} ap_i^{\alpha} |h_i|^2 + N_0	 \Big) \bigg).
\end{aligned}
\end{equation}
Then, the optimal solution of (\ref{P9}) can be obtained by iteratively optimizing $\boldsymbol{p}$, $\boldsymbol{\gamma}$, and $\boldsymbol{y}$ in the following problem:
\begin{subequations}\label{P10}
	\begin{align}
	\ \ \ & \underset{\boldsymbol{p}, \boldsymbol{\gamma}, \boldsymbol{y}}{\text{maximize}}\ \ \ \ \ \ \ 	F_q(\boldsymbol{p}, \boldsymbol{\gamma}, \boldsymbol{y})\\
	& \text{subject to} \ \ \ \ \ \ \text{(\ref{P9}b), (\ref{P9}c)}.
	\end{align}
\end{subequations}
The detailed procedures are as follows:
%The FP-based method (as shown in Algorithm \ref{al3}) solves (\ref{P10}) by alternately optimizing a) the transmit power $\boldsymbol{p}$ and b) auxiliary variables $\boldsymbol{\gamma},\boldsymbol{y}$, detailed as below.
\begin{enumerate}[a)]
\item Optimize power allocation $\boldsymbol{p}$ with fixed $\boldsymbol{\gamma}$ and $\boldsymbol{y}$: When $\boldsymbol{\gamma}$ and $\boldsymbol{y}$ are fixed, problem (\ref{P10}) reduces to a convex problem. The optimal power allocation $p_k^\ast, \forall k$, can be obtained via standard convex optimization methods, e.g., interior point method \cite{boydConvex2004}.
\item Update variables $\boldsymbol{\gamma}$ and $\boldsymbol{y}$: With the obtained $p_k^\ast$, we update $\gamma_{k}$ as
\begin{equation}\label{y3}
\gamma_k^\ast = \frac{p_k^\ast|h_k|^2}{\sum_{i=k+1}^{K}p_i |h_i|^2 + \sum_{i=1}^{K} ap_i ^{\alpha} |h_i|^2 + N_0},\forall k.
\end{equation}
Besides, we update $y_k$ by solving $\frac{\partial F_q}{\partial y_k }=0$ as
\begin{equation}\label{y4}
y_k^* = \frac{\sqrt{\omega_k (\gamma_k+1)|h_k|^2 p_k^\ast}}{\sum_{i=k}^{K}|h_i|^2 p_i +\sum_{i=}^{K} ap_i^{\alpha} |h_i|^2  + N_0}, \forall k.
\end{equation}
\end{enumerate}
Procedures a) and b) operate alternatively and finally stop when $|F_q|$ is smaller than a prescribed threshold $\epsilon$. We summarize the procedures to solve (\ref{P10}) in Algorithm \ref{al3}.
%\emph{a) Optimizing power allocation $\boldsymbol{p}$ with fixed $\boldsymbol{\gamma}$ and $\boldsymbol{y}$:} When $\boldsymbol{\gamma}$ and $\boldsymbol{y}$ are fixed, problem (\ref{P10}) is a convex problem. The optimal power allocation $p_k^\ast, \forall k$, can be obtained via standard convex optimization methods, e.g., interior point method \cite{boydConvex2004}.
%
%\emph{b) Updating variables $\boldsymbol{\gamma}$ and $\boldsymbol{y}$:} With the obtained $p_k^\ast$, we update $\gamma_{k}$ as
%\begin{equation}\label{y3}
%\gamma_k^\ast = \frac{p_k^\ast|h_k|^2}{\sum_{i=1}^{k-1}p_i |h_i|^2 + \sum_{i=1}^{K} ap_i ^{\alpha} |h_i|^2 + N_0},\forall k.
%\end{equation}
%that is, given power allocation $\boldsymbol{p}$, the optimal $\boldsymbol{\gamma}$ can be explicitly determined by setting $\frac{\partial F_l}{\partial \gamma_k }$ to zero.
\vspace{-10pt}
%\vspace{6pt}
\begin{algorithm}
	%	\small
	%		\scriptsize
	\SetAlgoLined
	\SetKwData{Left}{left}\SetKwData{This}{this}\SetKwData{Up}{up}
	\SetKwRepeat{doWhile}{do}{while}
	\SetKwFunction{Union}{Union}\SetKwFunction{FindCompress}{FindCompress}
	\SetKwInOut{Input}{input}\SetKwInOut{Output}{output}

	\textbf{Initialization:} Initialize $ \boldsymbol{p}, \boldsymbol{y}, \boldsymbol{\gamma}$ to a feasible value, $\text{iter}=0$, $N_{iter}$ as the number of maximum iterations;\\
	\Repeat{$|F_q| \textless \epsilon$  {\rm or iter = }$ N_{iter}$} {
			Update $\boldsymbol{p}$ by solving (\ref{P10}) with given fixed  $\boldsymbol{y},\boldsymbol{\gamma}$;\\
    	Update the auxiliary variable $\boldsymbol{\gamma}$ by (\ref{y3});\\
		Update the auxiliary variable $\boldsymbol{y}$ by (\ref{y4});\\
	$	\text{iter} = \text{iter}+1$ (\ref{y4});\\
	}
	\textbf{return} power allocation solution $\boldsymbol{p}^*.$
	\caption{The procedures to solve (P3)}
	\label{al3}
\end{algorithm}
\vspace{-20pt}
\subsection{Special Case of Equal Weighting Factors} \label{sec4.a}
To gain some insight of the optimal power control solutions, we study a special case of (\ref{P8}) where users have equal weighting factors. In other words, we consider the following  the sum rate maximization problem
\begin{subequations}\label{P4}
	\begin{align}
	\text{(P4)}:\ \ & \underset{\boldsymbol{p}}{\text{maximize}}\ \ \ \ \ \ \sum_{k=1}^{K} R_k \\
	& \text{subject to}\ \ \ \ \ \ \text{(\ref{P9}b), (\ref{P9}c)} .
	\end{align}
\end{subequations}
As shown in \eqref{s1}, the sum rate of a two-user uplink NOMA system is irrelevant to the decoding order. This result holds for the multi-user case following the similar steps in \eqref{s1}. Accordingly, (\ref{P4}a) is calculated as
\begin{equation}
\begin{split}
\begin{aligned}
\sum_{k=1}^{K} R_k =\log_2 \left(1+ \frac{ \sum_{k=1}^{K}p_k|h_k|^{2}}{ \sum_{k=1}^{K}a p_k^\alpha |h_k| ^{2} + N_0} \right).
\end{aligned}
\end{split}
\end{equation}
%which we using the fact that the sum rate is irrelevant to the decoding order $\pi$.
%% and $p_i$ denotes the power allocation of $UE$ in $i$th link.
%Note that the objective function has a similar form comparing to (P1), we can also applying Dinkelbach method for solving problem (P4). As a result, (P4) can be expressed as the equivalent optimization problem below
Therefore, (P4) is equivalent to
\begin{subequations}\label{P7}
	\begin{align}
	 & \underset{\boldsymbol{p}}{\text{maximize}}\ \ \ \ \ \  \frac{ \sum_{k=1}^{K}p_k|h_k|^{2}}{ \sum_{k=1}^{K}a p_k^\alpha |h_k| ^{2} + N_0} \\
	& \text{subject to}\ \ \ \ \ \ \text{(\ref{P9}b), (\ref{P9}c)},
	\end{align}
\end{subequations}
%
%\begin{equation}\label{P12}
%\underset{\boldsymbol{p}}{\text{maximize}} \ \frac{ \sum_{k=1}^{K}p_k|h_k|^{2}}{ \sum_{k=1}^{K}a p_k^\alpha |h_k| ^{2} + N_0}, \text{subject to \ (\ref{P9}b), (\ref{P9}c)},
%\end{equation}
which can be solved by using the Dinkelbach method. In particular, we introduce an auxiliary variable $\hat{\gamma}$ and reformulate (\ref{P7}) as
\begin{subequations}\label{P6}
	\begin{align}
\ \ & \underset{\hat{\gamma}, \boldsymbol{p}}{\text{maximize}}\ \ \ F_1 = \sum_{k=1}^{K}p_k|h_k|^{2} - \hat{\gamma}\left( \sum_{k=1}^{K}a p_k^\alpha |h_k|^{2} + N_0 \right)   \\
	& \text{subject to}\ \ \ \ \ \ \text{(\ref{P9}b), (\ref{P9}c)}.
	\end{align}
\end{subequations}
The optimal solution of (\ref{P7}) can be obtained by iteratively optimizing $\boldsymbol{p}$ and $\hat{\gamma}$ in (\ref{P6}) following similar procedures in Algorithm 1. The details are omitted here for brevity.
%The detailed procedures are as follows:
%\emph{a) Optimizing power allocation $\boldsymbol{p}$ with fixed $\hat{\gamma}$:}
%In this case, (\ref{P6}) is a convex problem that can be efficiently solved by the standard convex optimization methods.

To examine the property of the optimal power allocation policy, we introduce the partial Lagrangian function of problem (\ref{P6}) as
\begin{equation}
\begin{aligned}
\mathcal{L}=& F_1- \sum_{k=1}^{K}\lambda_k \Bigg(p_k|h_k|^2- \big(2^{r_k} -1 \big) \cdot \\
& \bigg( \sum_{i=k+1}^{K} p_i |h_i|^2 + \sum_{i=1}^{K} ap_i^{\alpha} |h_i|^2 + N_0 \bigg) \Bigg) ,
\end{aligned}
\end{equation}
where $\lambda_k$'s denote the non-negative Lagrange multiplier associated with the corresponding constraints (\ref{P9}b).
By taking the derivative of $\mathcal{L}$ with respect to $p_k$ and set it to 0, we have
\begin{equation}\label{la}
\begin{split}
\frac{\partial \mathcal{L} }{\partial p_k} &= \hat{\gamma} a p_k^{\alpha-1}\alpha|h_{k}|^{2} -|h_{k}|^{2}-  \\ &\left(\sum_{i=1}^{K} \lambda_i\left( 2^{r_i}-1\right)\right) \left(a p_k^{\alpha -1} \alpha |h_{k}|^{2}\right)  \\ & + \left( \sum_{i = k+1}^{K} \lambda_i\left( 2^{r_i}-1\right)\right) |h_{k}|^{2} - \lambda_k |h_{k}|^{2} = 0.
\end{split}
\end{equation}
Then, the optimal transmit power (denoted as $p_k^\ast$) can be obtained from \eqref{la} as
%By solving $\frac{\partial\mathcal{L}}{\partial p_k}=0$, we obtain the optimal  power $p_k^\ast$ as
\begin{equation}\label{eq2}
p_k^* =  \left[\sqrt[\alpha-1]{\frac{1+\lambda_k -\sum_{i = k+1}^{K} \lambda_i\left( 2^{r_i}-1\right)}{\left(  \hat{\gamma} + \sum_{i = 1}^{K} \lambda_i(2^{r_i}-1)  \right) a\alpha }}\right]_0^{\hat{p}_k},
\end{equation}
where $[ \cdot ]^{y}_{x}=\min(\max (\cdot ,x),y)$. With \eqref{eq2}, we can update $\lambda_k's$ using sub-gradient or ellipsoid method \cite{boydConvex2004}, until convergence condition is met. When $r_k=0$, $\forall k$, problem \eqref{P4} is simplified to the sum rate maximization problem without any rate constraint, such as the two-user case in \eqref{sum rate_problem}. In this case,  we have $\lambda_k=0$ and
\begin{equation}\label{eq3}
p_k^* = \left[\sqrt[\alpha-1]{\frac{1}{\hat{\gamma} a\alpha}}\right]_0^{\hat{p}_k}.
\end{equation}
This implies that in the absence of individual rate constraints, users transmit data using equal power to maximize the sum rate, provided that $\hat{p_k}\geq \sqrt[\alpha-1]{\frac{1}{\hat{\gamma} a\alpha}}, \forall k $. If not, the optimal transmit power of each user varies depending on their maximum transmit power $\hat{p_k} = \min\left[\bar{p_k},p_k^{\text{opt}} \right]$.

To summarize the procedure, given $\lambda_k's$ and an initialized $\hat{\gamma}$, the optimal power of maximizing $\mathcal{L}$ is given by \eqref{eq2}. After computing the primary variables $\boldsymbol{p}$, we employ the sub-gradient or ellipsoid method to update the Lagrange multipliers $\lambda_k's$, leading us towards the optimal solution $p^{*}_{k},\forall k$. Having obtained $p_k^\ast,\forall k$, we update $\hat{\gamma}$ as
\begin{equation}\label{gamma2}
\hat{\gamma}^* =  \frac{ \sum_{k=1}^{K}p^\ast_k|h_k|^{2}}{ \sum_{k=1}^{K}a p_k^{\ast \alpha} |h_k| ^{2} + N_0},
\end{equation}
and $F_1$ as \eqref{P6}. Then, the iteration proceeds until  $|F_1|\textless \epsilon$, where we reach the optimal solution $\{p^*_k\}'s $ to \eqref{P4}.
\section{Extension to Broadband Channel}\label{section4}
In this section, we demonstrate how the proposed PA distortion model can be extended to accommodate broadband communication systems with dissimilar OFDM sub-channel coefficients.
The block diagram of broadband OFDM system with nonlinear PA is depicted in Fig. \ref{ofdm}. Let $\boldsymbol{\tilde{s}}\in \mathcal{C}^{N\times 1}$ denote the baseband symbols to be transmitted on the $N$ sub-carriers. These symbols are fed into the OFDM modulation module, which consists of a serial-to-parallel converter, IFFT calculation, cyclic prefix insertion, and a parallel-to-serial converter. The output of the OFDM module is
\begin{equation}
s_n = \frac{1}{N}\sum_{m=0}^{N-1} \tilde{s}_m e^{j2\pi mn/N} , 0 \leq n \leq N+N_g-1,
\end{equation}
where $N_g$ denotes the length of cyclic prefix.
The sequence $\boldsymbol{s}=(s_0,\cdots s_{N+N_g-1})$ is then upsampled using a proper interpolation filter to generate $\boldsymbol{u}$ \cite{chiuPredistorter2008a}, which is then fed into the DPD-PA system.
\begin{figure}[tbp]
	\centering
	%	\vspace{-8pt}
	\includegraphics[scale=0.6]{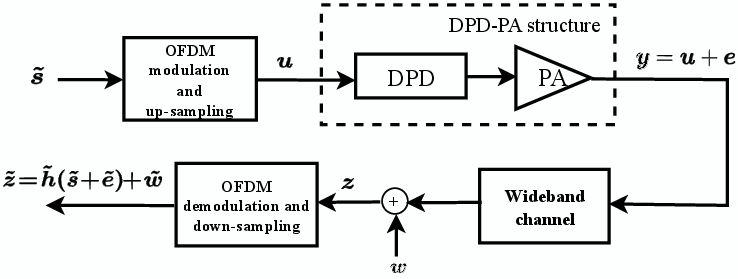}
	\captionsetup{font=footnotesize}
	\caption{Baseband signal passes by OFDM system with nonlinear PA in transmitter.}
	%	\vspace{-1pt}
	\label{ofdm}
\end{figure}
The PA output signal $y$ is modeled in \eqref{bussgang} and transmitted by the antenna.

At the receiver end, the cyclic prefix is removed, the signal is downconverted to baseband, and subsequently passed through a serial-to-parallel converter and an FFT module.
The baseband representation of the received signal on $m$th sub-carrier can be expressed as
\begin{equation}
\tilde{z}_m = \tilde{h}_m(\tilde{s}_m+\tilde{e}_m)+\tilde{w}_m, \forall m = 0,...N-1,
\end{equation}
where $\tilde{z}_m$ denotes the $m$th component of $\boldsymbol{\tilde{z}}$. We denote $\boldsymbol{\tilde{z}}=\boldsymbol{Uz}$, $\boldsymbol{\tilde{s}}=\boldsymbol{Us}$, $\boldsymbol{\tilde{e}}=\boldsymbol{Ue}$, and $\boldsymbol{\tilde{w}}=\boldsymbol{Uw}$ as the DFT of the random vector $ \boldsymbol{z,s,e}$, and $ \boldsymbol{w}$, respectively. Here, $\boldsymbol{U}$ is the unitary matrix and
\begin{equation}
\boldsymbol{U}_{mn} = \frac{1}{N} e^{(-j2\pi nm)/N}, m,n  =0,...,N-1.
\end{equation}
It is known that if $\boldsymbol{\tilde{s}}$ is an i.i.d. complex Gaussian vector, the distribution of $ \boldsymbol{U\tilde{s}} $ is identical to $\boldsymbol{\tilde{s}}$ for every unitary matrix $\boldsymbol{U}$.
Therefore, we have $\tilde{s}_m \!\!\sim\! \mathcal{CN}(0,P_{T}^m)$, $\tilde{w}_m \!\!\sim\! \mathcal{CN}(0,\frac{B}{N}N_1)$, and $\tilde{e}_m \!\!\sim\! \mathcal{CN}(0,\frac{P_N}{N})$ as the $m$th component of $\boldsymbol{\tilde{s},\tilde{w}}$, and $\boldsymbol{\tilde{e}}$, respectively. Here, $B$ denotes the total bandwidth, $N_1$ denotes the power spectral density of the receiver noise,
$ P_{T}^m$ denotes the transmit power of the $m$th sub-carrier with $ \sum_{m=0}^{N-1} P_{T}^m=P_T$, and $P_N = aP_T^{\alpha}$ denotes the total PA distortion power.
Accordingly, the communication rate of the point-to-point link on the $m$th sub-carrier is
\begin{equation}\label{OFDMrate}\small
R_m =  \frac{B}{N}\log_2 \left( 1 + \frac{P_T^m |\tilde{h}_m|^2}{\frac{a P_T^\alpha}{N} |\tilde{h}_m|^2 + \frac{B}{N} N_1 }\right), \forall m.
\end{equation}
In the subsequent steps, we can perform a similar rate analysis as we did in Sections \ref{section2} and \ref{section3} by taking into account the data transmitted on parallel sub-channels.
Furthermore, when the channel gain and allocated power on $m$th are identical for all sub-carriers, i.e., $\tilde{h}_m = \tilde{h}$ and $P_T^m = \frac{P_T}{N}$, \eqref{OFDMrate} reduces to
	\vspace{4pt}
\begin{equation}
R_m = \frac{B}{N}\log_2 \left(1 + \frac{P_T |\tilde{h}|^2}{a P_T^\alpha |\tilde{h}|^2 +B N_1 }\right),\forall m.
\end{equation}
We observe that the sum rate $\sum_{m=0}^{N-1}R_m$ has the same form of the point-to-point baseband signal model in \eqref{Ri}.

\section{Numerical Results}\label{section5}
In this section, we evaluate the performance of the uplink NOMA system under consideration via numerical simulations. Unless otherwise stated, we consider $K=4$ UEs with identical non-linear PAs. The distance between the $k$th UE and the BS is $d_k =  60+ 20(k-1)$ meters. We consider a path-loss channel between the $k$th UE and the BS, where the channel gain is $|h_k|^2= G_A \left(\frac{3\times10^8}{4\pi f_c d}\right)^{\sigma}$. Here, $f_c=2.4$ GHz is the carrier frequency. $\sigma=2.6$ is the path-loss exponent. $G_A = 4.11$ denotes the antenna gain.
The channel noise power at the BS is $N_0=\sigma^2B$, where $\sigma^2=-174$ dBm/Hz is the power spectral density of AWGN and $B=30$ MHz is the operating bandwidth of UEs.
Besides, we set the maximum transmit power at UEs as $\bar{p}= 36$ dBm. The default PA distortion parameters are $(a, \alpha)=(0.0032,1.3552)$, corresponding to Fig. \ref{NMSE}(b). The stopping criterion in Algorithm \ref{al3} is set as $\epsilon = 10^{-4}$. To demonstrate the importance of accurate PA distortion modeling to system throughput and resource
allocation efficiency, in the following, we first study the capacity region of a two-user uplink NOMA system under NDM and IDEAL PA model, respectively. Then, we investigate weighted sum rate performance of multi-user uplink NOMA systems.
For convenience, we denote the optimal power control policies as
\begin{itemize}
	\item PC-NDM: This assumes the proposed non-linear PA model and follows the optimal power control method in Algorithm \ref{al3};
	\item PC-IDEAL: This assumes an ideal PA without nonlinear distortion by setting $a = 0$ in the PA distortion model \eqref{PN}, and optimizes the transit power using Algorithm \ref{al3}.
\end{itemize}
%[加个reference].

\subsection{Capacity Region of Two-user NOMA}
Fig. \ref{cp} presents the capacity region of a two-user uplink NOMA system, comparing the performance of non-linear distortion power model (NDM) and the ideal PA (IDEAL). In this simulation, we set the distances of the two UEs to the BS as $d_1 = 120$ and $d_2 = 80$ meters, respectively.
As shown in the figure, the presence of non-linear PA distortion significantly shrinks the achievable region compared to the ideal case without distortion. Meanwhile, we have plotted two operating points that maximize the sum rate following the PC-NDM power control (point G) and the PC-IDEAL power control method (point H) model. The results indicate that the optimal power allocation, armed with the knowledge of the NDM, offers 10.15\%  higher throughput than when assuming an ideal PA. It is evident that the operating point H, which assumes an inaccurate PA model, deviates from the capacity boundary, leading to inefficient resource usage.
%To further demonstrate the importance of accurate PA distortion modeling
%to the system throughput and resource allocation efficiency, in the following, we study the sum rate and weighted sum rate performance of the considered NOMA system by assuming NDM and ideal PA model, respectively. For convenience, we denote the optimal power control policy under NDM and ideal PA as
%PC-NDM and PC-IDEAL, respectively.

\begin{figure}[tbp]
	\centering
	\vspace{6pt}
	\includegraphics[scale=0.6]{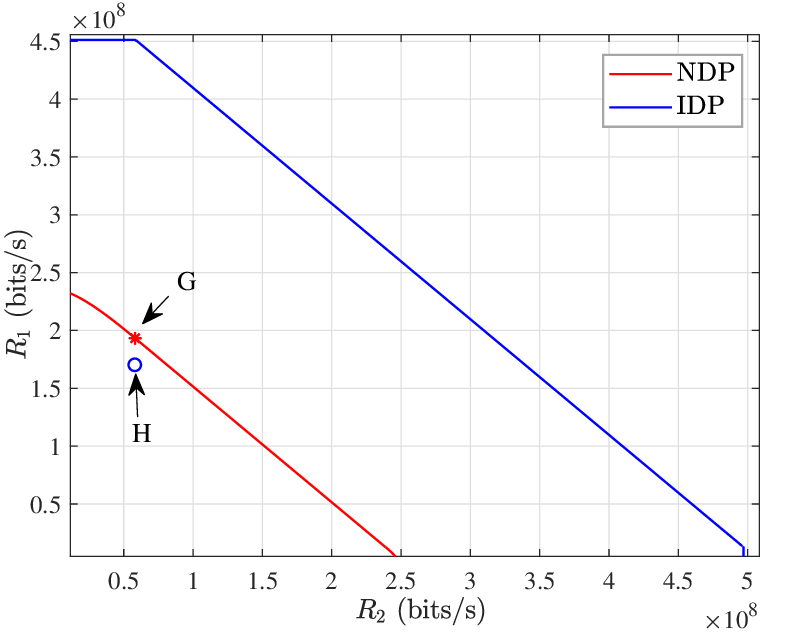}
	\captionsetup{font=footnotesize}
	\caption{Capacity region of two-user uplink NOMA system, where the operating point G (PC-NDM) is located at $ (R^*_1,R^*_2) \!\!=\!\! (1.933, 0.582)\!\!\times\!\!10^8 $ bits/s with the optimal transmit power $p_1^\ast \!= \!p_2^\ast\! = \!$ 25.9 dBm, and point H (PC-IDEAL)  at $(R_1,R_2) \!\!=\!\!(1.703 , 0.579)\!\!\times \!\!10^8$ bits/s with transmit power 36.0 dBm.}
	%	\vspace{-1pt}
	\label{cp}
\end{figure}
\subsection{Throughput Performance Comparison}
\begin{figure}[tbp]
	\centering
	\vspace{8pt}
	\includegraphics[scale=0.6]{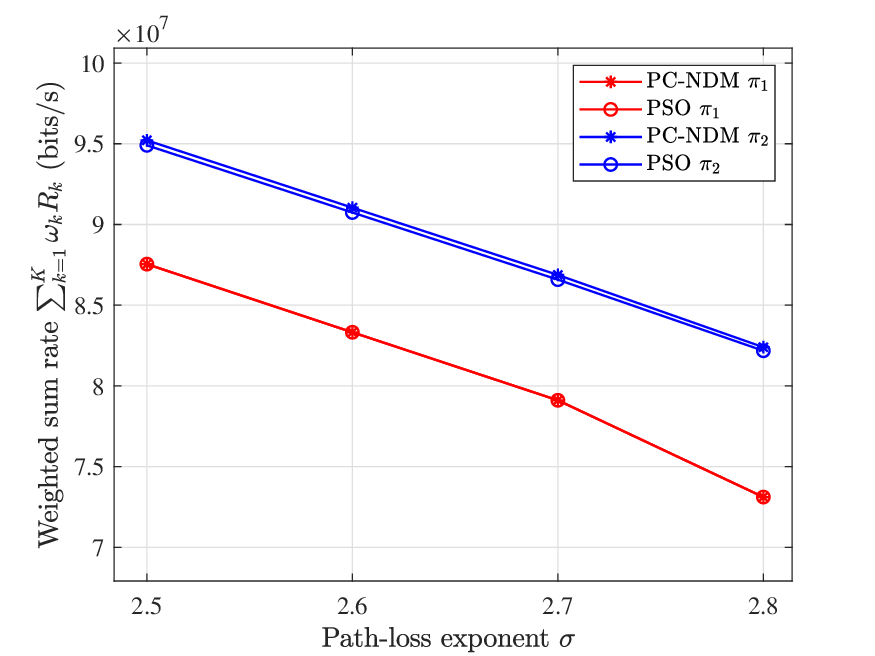}
	\captionsetup{font=footnotesize}
	\caption{Weighted sum rate vs. the path-loss exponent $\sigma$. }
	%	\vspace{-1pt}
	\label{PSO}
\end{figure}
In Fig. \ref{PSO}, we compare the WSR performance of the proposed algorithm and a benchmark particle swarm optimization (PSO) algorithm under different path-loss exponents  \cite{guoPower2018}. PSO is a meta-heuristic algorithm that iterates through the solution space in a structured manner, converging to an optimal solution after a sufficient number of iterations. The comparison shows that the two curves are on top of each other for all the cases considered. This demonstrates the effectiveness of the proposed WSR algorithm, as it performs comparably to the PSO algorithm, a well-established optimization technique.

In Fig. \ref{L2}, we examine the sum rate performance when no fairness constraint is enforced ($r_k=0,\forall k$ in (P4)). Under all considered path-loss exponents $\sigma$'s, PC-NDM outperforms PC-IDEAL by an average of 13.54\% ($B = $ 30 MHz) and 15.48\% ($B = $ 20 MHz) higher sum rate, respectively. This is because PC-IDEAL power control policy causes severe distortion noise. Since the BS receives larger PA distortion noise under stronger communication links, the performance advantage of PC-NDM over PC-IDEAL is more significant at small $\sigma$'s.

\begin{figure}[tbp]
	\centering
	\vspace{10pt}
	\includegraphics[scale=0.6]{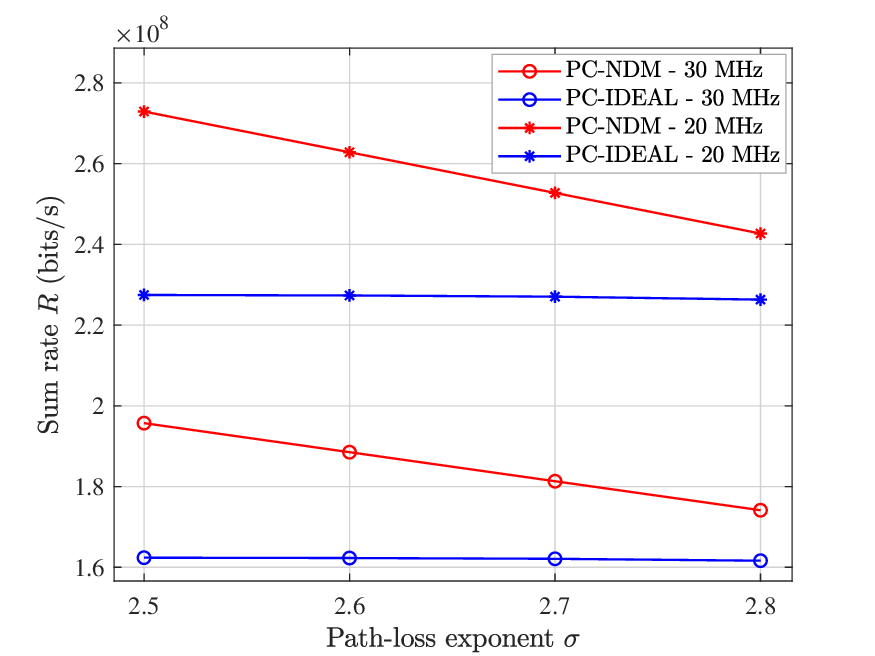}
	\captionsetup{font=footnotesize}
	\caption{Sum rate $R$ vs. the path-loss exponent $\sigma$.}
	\label{L2}
\end{figure}
\begin{figure}[tbp]	
	\centering
	\vspace{2pt}
	\includegraphics[scale=0.6]{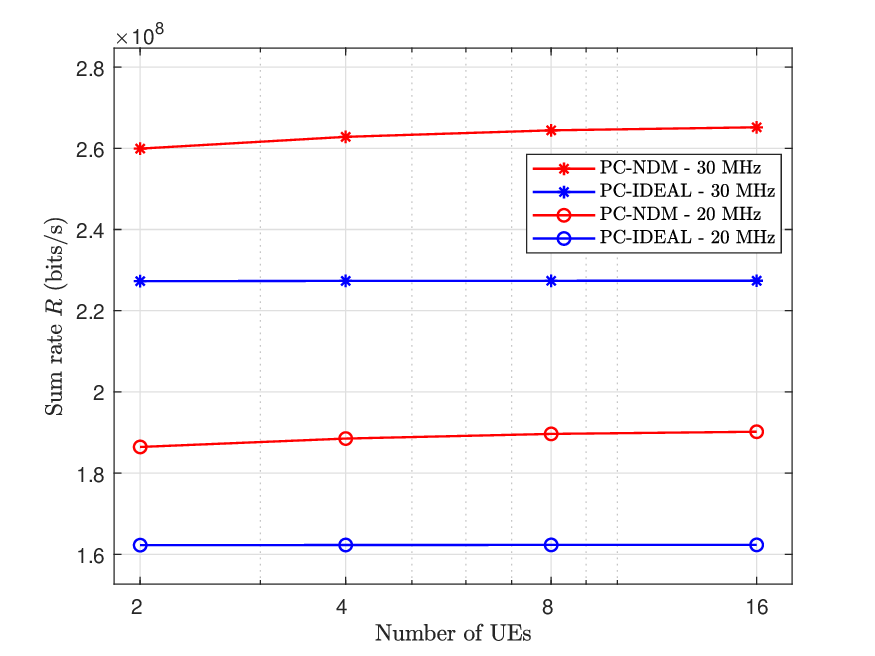}
		\captionsetup{font=footnotesize}
	\caption{Sum rate $R$ vs. number of UEs}
	%	\vspace{-1pt}
	\label{number sum rate}
\end{figure}

In Fig. \ref{number sum rate}, we evaluate the sum rate performance as the number of UEs varies from 2 to 16. For the considered operating bandwidth 20 MHz and 30 MHz, PC-NDM offers on average 16.32\% and 18.31\% higher sum rate over PC-IDEAL, respectively.
As the number of UEs grows, the sum rate of PC-NDM increases while that of PC-IDEAL remains constant due to the severe aggregate distortion power caused by more transmitting UEs.
Essentially, this is because the nonlinear distortions cannot be canceled by SIC at  the receiver.

\begin{figure}[tbp]
	\centering
	\vspace{2pt}
	\includegraphics[scale=0.6]{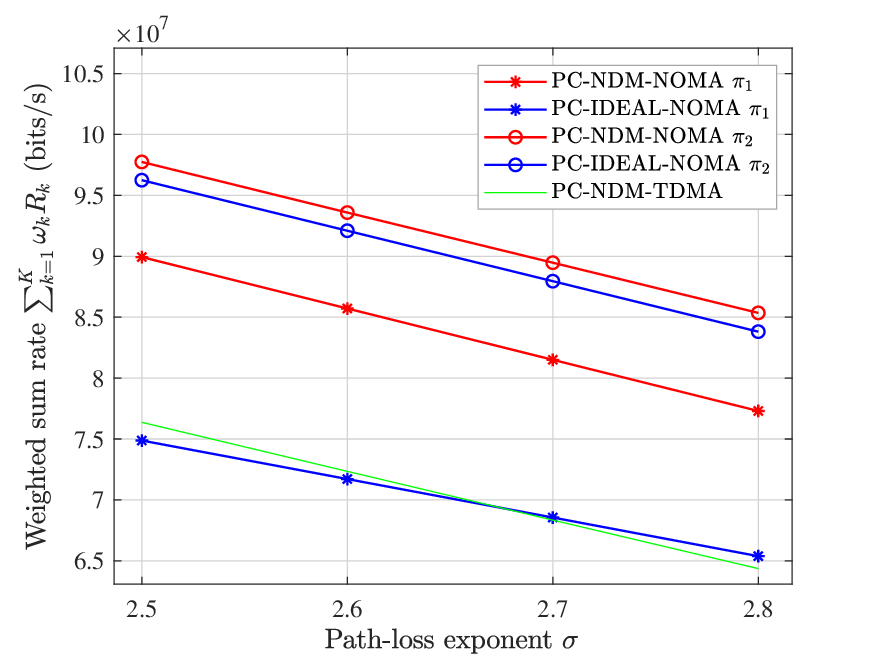}
	\captionsetup{font=footnotesize}
	\caption{Weighted sum rate vs. the path-loss exponent $\sigma$.}
	%	\vspace{-1pt}
	\label{tdma}
\end{figure}

\begin{figure}[tbp]
	\centering
	\vspace{4pt}
	\includegraphics[scale=0.6]{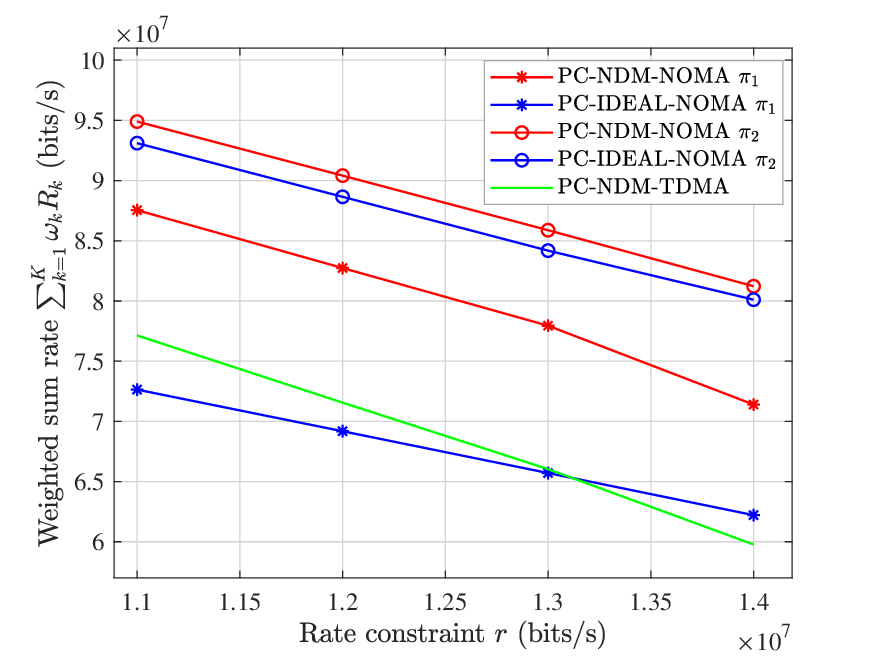}
	\captionsetup{font=footnotesize}
	\caption{Weighted sum rate vs. rate constraint $r_k = r,\forall k$.}
	%	\vspace{-1pt}
	\label{rate_constraint}
\end{figure}
In Fig. \ref{tdma} and Fig. \ref{rate_constraint}, we investigate the weighted sum rate performance of the considered system under varying path-loss exponents and rate constraints, respectively. Here, the weighting factors are set as $\omega=(0.1,0.2,0.3,0.4)$. We compare the performance of PC-NDM and PC-IDEAL under two different decoding orders, i.e., $\pi_{1}\!=\!4\!\rightarrow\!3\!\rightarrow\!2\!\rightarrow\!1$ and $\pi_{2}\!=\!1\!\rightarrow\!2\!\rightarrow\!3\!\rightarrow\!4$. To ensure the feasibility of PC-IDEAL, we conduct the following projection:
\begin{subequations}\label{projection}
	\begin{align}
	\underset{\mathcal{H}}\Pi (\boldsymbol{p}')=  & \arg \underset{\boldsymbol{p}} \min \ \ \ \Vert \boldsymbol{p} - \boldsymbol{p}' \Vert \\
	& \text{subject to}\ \  \text{(\ref{P9}b), (\ref{P9}c)},
	\end{align}
\end{subequations}
where $p\prime$ is the power control solution of PC-IDEAL. $\mathcal{H}$ is the feasible region of the WSRMax problem \eqref{P10}.
Besides, we also plot the corresponding weighted sum rate performance of  traditional time duplexing multiple access system (labeled as PC-NDM TDMA).
In both figures and under both decoding orders, the proposed power control policy achieves higher data rate. The performance advantage is more evident under decoding order $\pi_1$. Interestingly, Fig. \ref{rate_constraint} shows that even simple TDMA achieves higher data rate than NOMA under the PC-IDEAL power control policy when the rate constraint $ r \leq 1.3 \times 10^7$. This is due to the severe PA distortion interference caused by the improper power control.
These findings confirm the importance of accurate PA distortion modeling. Without accurate distortion modeling, even traditional methods like TDMA can outperform NOMA, underscoring the significant role of PA distortion modeling in maximizing system performance.

%%\subsection{Capacity Limits}
%The capacity with ideal hardware grows without bound as $N \rightarrow \infty$, while the lower bounds converage to finite limits under PA nonlineartiy.
%The capacity bounded as
%\begin{equation}
%\lim\limits_{N \rightarrow \infty} \log_2 \left(1+ \frac{ \sum_{i=1}^{N}p_i|h_{i}|^{2}}{ \sum_{i=1}^{N}a p_i^\alpha |h_{i}|^{2} + \sigma_1^2} \right),
%\end{equation}
%which is equivalent to bounded as
%\begin{equation}
%	\lim\limits_{N \rightarrow \infty}  \frac{ \sum_{i=1}^{N}p_i|h_{i}|^{2}}{ \sum_{i=1}^{N}a p_i^\alpha |h_{i}|^{2} + \sigma_1^2},
%\end{equation}
%we assume that each $UE$ has a identical channel $h$, then we have
%\begin{equation}
%\lim\limits_{N \rightarrow \infty}  \frac{ N p|h|^2}{ N a p^\alpha |h|^2 + \sigma_1^2} = \frac{1}{a p ^{\alpha-1}}.
%\end{equation}
%Therefore, given a power allocation $p$, the uplink capacity bounded as
%\begin{equation}
%	\lim\limits_{N \rightarrow \infty}  \frac{ \sum_{i=1}^{N}p_i|h_{i}|^{2}}{ \sum_{i=1}^{N}a p_i^\alpha |h_{i}|^{2} + \sigma_1^2} = \log_2(1+ \frac{1}{a p ^{\alpha-1}}).
%\end{equation}

\section{CONCLUSION}\label{section6}
In this paper, we established a new PA distortion model based on real-world measurements, where the distortion noise power is a polynomial function of PA transmit power. Leveraging this distortion model, we accurately characterized the capacity region of a multi-user uplink NOMA. For practical engineering applications, we formulated a general WSR maximization problem and proposed an efficient power control algorithm to achieve optimal performance. Numerical results show that the distortion noise power significantly shrinks the achievable capacity region of NOMA. Meanwhile, the proposed power control method offers over 13\% data rate improvement in extensive experiments compared to the scenario where an ideal PA is assumed and a sub-optimal power control policy adopted. With these findings, we highlight the importance of accurate PA distortion modeling to the system throughput performance and resource allocation efficiency. Our research provides valuable insights for the design and optimization of future NOMA systems, particularly with regards to power control under real-world PA distortion conditions.

\appendices
\section{Proof of Proposition 1}  \label{app1}
\emph{Proof}: To prove Proposition 1, we need to show 1) the line $l^*_{sum}$ intersects with the boundary of both $R^{\pi_1}$ and $R^{\pi_2}$,  and 2) $l^*_{sum}$ does not enter the interior of the boundaries of $R^{\pi_1}$ and $R^{\pi_2}$.

The first argument can be easily proved because the optimal solution to (\ref{sum rate_problem}) is also a feasible solution to (P1) for some $\tau$, indicating that $ l_{sum}^* \bigcup R^{\pi_1}$ is non-empty. Similarly, we have $ l_{sum}^* \bigcup R^{\pi_2}$ is non-empty. To show the second argument,
we can use the method of proof by contradiction. Without loss of generality, suppose that $l_{sum}^*$ enters the interior of $R^{\pi_1}$. By definition, there exit a point $(R_1^{(1)},R_2^{(1)}) \in l_{sum}^*$ and a point $(R_1^{(2)},R_2^{(2)}) \in R^{\pi_1}$, such that at least one of the following two conditions holds strictly, a) $R_1^{(1)}\leq  R_1^{(2)}$ and  $R_2^{(1)}<R_2^{(2)}$; b) $R_1^{(1)}<  R_1^{(2)}$ and  $R_2^{(1)} \leq R_2^{(2)}$. In any case, we have $R_1^{(1)}+ R_2^{(1)} <R_1^{(2)}+R_2^{(2)}$. Because $(R_1^{(2)},R_2^{(2)})$ is also a feasible solution to (\ref{sum rate_problem}), this leads to a contradiction that the objective of (\ref{sum rate_problem}) can be further improved such that $(R_1^{(1)},R_2^{(1)}) \notin l_{sum}^*$. Therefore, we reach to the conclusion that $l_{sum}^*$ does not enter the interior of $R^{\pi_1}$. Similar argument can be made that $l_{sum}^*$ does not enter the interior of $R^{\pi_2}$ as well. This leads to the proof of argument 2) and also Proposition 1.
$\hfill \qedsymbol$

\bibliographystyle{IEEEtran}
\bibliography{IEEEabrv,cite1}

\end{document}